\newcommand{\tsecompldate}{Friday, April 22, 2005 at 12:30}
\documentclass[12pt]{article}

\usepackage{amssymb}

\usepackage{graphicx}

\newcommand{\vol}[1]{\textbf{#1}}

\newcommand{\tpaptitle}[1]{``#1'',}
\newcommand{\tpretitle}[1]{``#1'',}
\newcommand{\tarttitle}[1]{``#1'',}

\newcommand{\tbktitle}[1]{``#1''}

\newcommand{\tISBN}[1]{#1}

\newcommand{\tref}[1]{(\ref{#1})}

\newcommand{\tnotpre}[1]{#1}
\newcommand{\tversion}{Draft version 2}

\newcommand{\tpre}[1]{}
\newcommand{\tprenote}[1]{}

\newcommand{\tnote}[1]{} 
\newcommand{\tcomment}[1]{} 
\newcommand{\tcommentx}[1]{}

\newcommand{\href}[2]{#2}
\newcommand{\eprint}[1]{\texttt{#1}}

\newcommand{\tsedevelop}[1]{{}}

\makeatletter

\@addtoreset{equation}{section} \makeatother \renewcommand{\vol}[1]{{\bf #1}}

\renewcommand{\tpaptitle}[1]{\emph{#1},}

\renewcommand{\tarttitle}[1]{\emph{#1},}

\renewcommand{\tbktitle}[1]{``#1''}
\renewcommand{\tISBN}[1]{}

\renewcommand{\tpretitle}[1]{{\em #1},}

\renewcommand{\tnotpre}[1]{}
\renewcommand{\tpre}[1]{#1}
\renewcommand{\tprenote}[1]{\footnote{#1}}
\renewcommand{\tversion}{Preprint version}

\newcommand{\half}{\frac{1}{2}}
\newcommand{\bea}{\begin{eqnarray}}
\newcommand{\eea}{\end{eqnarray}}
\newcommand{\beq}{\begin{equation}}
\newcommand{\eeq}{\end{equation}}
\newcommand{\nnel}{\nonumber \\ {}}

\newcommand{\ra}{\rightarrow}

\begin{document}

\renewcommand{\thefootnote}{\fnsymbol{footnote}}

 \tpre{\begin{flushright}
 \tsecompldate \\
 \texttt{Imperial/TP/041102} \\
 \eprint{cond-mat/0411390} \\
 \tversion \\
 \tsedevelop{(\texttt{wwalkcondmat2.tex}  LaTeX-ed on \today ) \\}
 \end{flushright}
 \vspace*{1cm} }

\begin{center}
{\Large\textbf{Scale Free Networks from
Self-Organisation}}\tnote{tnotes such as this not present in final
version} \\
\tpre{\vspace*{1cm} }
 {\large T.S.\ Evans\footnote{email:
\texttt{T.Evans}\emph{@}{ic.ac.uk}\tnotpre{, \tsecompldate} } }
 \\
 \tpre{\vspace*{0.5cm}}
 Theoretical Physics,
 Blackett Laboratory, Imperial College London,\\
 Prince Consort Road, London, SW7 2BW,  U.K.
\\
 \tpre{\vspace*{0.5cm}}
 {\large J.P. Saram\"aki}\footnote{email:
 \texttt{jsaramak}\emph{@}{lce.hut.fi}}
 \\
 \tpre{\vspace*{0.5cm}}
 Laboratory of Computational Engineering,
 Helsinki University of Technology, \\ P.O.~Box 9203,
 FIN-02015 HUT, Finland
\end{center}



\begin{abstract}
We show how scale-free degree distributions can emerge naturally
from growing networks by using random walks for
selecting vertices for attachment. This result holds for several
variants of the walk algorithm and for a wide range of parameters.
The growth mechanism is based on using local graph information only,
so this is a process of self-organisation. The standard mean-field
equations are an excellent approximation for network growth using
these rules. We discuss the effects of finite size on the degree
distribution, and compare analytical results to simulated networks.
Finally, we generalise the random walk algorithm to produce weighted
networks with power-law distributions of both weight and degree.
\end{abstract}

\renewcommand{\thefootnote}{\arabic{footnote}}
\setcounter{footnote}{0}


\section{Introduction}

Many networks seen in the real world have a degree distribution
which is a power-law for large degrees
\cite{AB,DM01,Watts03,Newman03,TSE04}, at least to some
approximation. This means that there are many more vertices with
large degrees, `hubs' of a network, than one would find with the
traditional Erd\H{o}s and R\'enyi random graphs with their
short-tailed Poisson degree distribution \cite{ER59}. Such long
tailed distributions have been of considerable interest for some
time in a wide range of fields, see \cite{Mit04} for a brief
overview.\tnote{Some of the best known discussions are those of
Pareto 1897, Yule 1925, Lotka 1926, Zipf 1945, and the arguments
between Mandelbrot and Simon in the 1950's, but there are many
other examples.}

On the theoretical side, scale-free graphs are generated in
several models.  Most are characterised by a probability, $\Pi$,
for choosing a particular existing vertex in an existing graph to
which a new edge is to be added. In particular, if a finite
fraction of new edges are attached with probability proportional
to the degree $k$ of the existing vertices, $\Pi(k) \propto k$, at
least for large degree vertices, then the graph will be scale-free
\cite{AB,DM01,Watts03,TSE04,Mit04}. Such attachment of edges with
probability proportional to degree of target vertices is often
termed preferential attachment\footnote{Such a rich get richer
algorithm echoes the well known Pareto 80:20 law of economics. It
does not matter if the graph is growing, or if it is just being
rewired with fixed numbers of edges and vertices, or anything in
between.  If preferential attachment dominates for edge attachment
to large degree vertices, a scale-free graph will emerge for large
graphs.}. This is a feature of the model by Simon \cite{Simon55}
and of the more recent Barab\'{a}si and Albert model \cite{BA}.

However, a key result is that if the $\Pi(k) \propto k^\alpha$,
then for \emph{any}  $\alpha \neq 1$ we do not get a simple power
law degree distribution for large degree in the large graph limit
\cite{KRL}. So, if scale-free laws are often found in nature,
where does the precisely linear preferential attachment with
$\alpha=1$ come from?  Further, it is crucial to know what the
total number of edges is in a network to provide the normalisation
for the linear preferential attachment probability.  This is
simple for numerical models and theoretical analysis.  However, it
is a piece of \emph{global} information not usually available at
nodes in real systems.  The authors of web pages do not know, nor
do they care, how big the web is for instance.

It is evident that the processes shaping networks in the real
world are usually \emph{local}, i.e.\ they rely mostly on
structural properties of the networks in the neighbourhood of a
vertex. Hence, realistic models of network evolution should
likewise be based on local rules~\cite{Vaz00,Vaz02,KR02,Chung02}.
Here, our focus is on random walks on networks \cite{SK04,Kla}. A
random walk on a graph tends to arrive at a vertex with a
probability proportional to the number of ways of arriving at that
vertex, i.e.\ the degree of that vertex. A random walk can be
viewed as natural way for preferential attachment to appear using
only the local properties of a graph. For instance, consider the
graph of vertices representing film actors, joined if they have
appeared in the same film \cite{Watts03,TSE04}. One can imagine a
new actor has one or two initial contacts with established actors.
They may not know of any suitable jobs for the newcomer, but they
pass the word on to their contacts. These in turn might pass the
word on to their contacts, until by chance a suitable job is
found.  A new edge is formed to an existing node chosen by a walk
along existing links in the network and this is equivalent to
choosing a vertex proportional its degree. Indeed, in anthropology
it has long been noted that providing access to a wider pool of
resources than is locally available is often an important role of
many kinship networks.

The random walk algorithm illustrates how the network structure
can be driven naturally to a scale-free form as result of purely
local microscopic processes. It is the very structure of the graph
itself which guides the search, and thus it is not too surprising
that the asymptotic limit has a common feature, a scale-free
distribution. Although the algorithm itself is an idealisation, we
argue that the scale-free nature of many real world networks is a
consequence of network evolution driven by this type of mechanism.
For this argument to hold, the details of the random walk
mechanism should not change the outcome, i.e.\ the form of the
resulting distributions should be robust to variations in the
algorithm.

The purpose of this paper is to extend the work of Saram\"aki and
Kaski \cite{SK04} and to demonstrate the robustness of the walk
algorithm. First, we will discuss the mean-field equations for the
network evolution, the length scales present in finite-sized
networks, and the form of the degree distribution for finite-size
networks based on preferential attachment growth. Then, we will
present the generalised random walk algorithm, and compare results
from numerical simulations to theoretical ones. Finally, we will
generalise the algorithm of \cite{SK04} to the case of weighted
graphs, yielding asymptotically scale-free distributions of both
degree and weight.

\section{Mean Field Equations}

The mean field equations are a good approximation for the
behaviour of degree distributions in many different algorithms.
These will serve to fix our notation, but solutions to these
approximate equations also match practical models and we will be
referring to them later.

Consider a sequence of graphs $\{ G(t) \}$, consisting of $N(t)$
vertices and $E(t)$ edges.  Here $t$ is a \emph{time-like} integer
parameter, where in going from $t$ to $t+1$ we add a vertex a
fraction $\epsilon$ of the time, while each time adding on average
a total of $m$ edges\footnote{Note that for a realistic model $t$
is probably a monotonic function of the real physical time since
one might expect large graphs to grow faster in real time than
small ones. However all we require for our analysis is that the
number of edges added per new vertex is constant and this in turn
provides a definition of our $t$ parameter in terms of the growth
of any real world network.}. The total number of vertices, $N(t)$,
and the total number of edges, $E(t)$, grow on average as
\begin{eqnarray}\label{Nevolv}
  N(t) &=& \sum_k n(k,t) = N_0 + \epsilon t
\\
\label{Eevolv}
  E(t) &=& \half \sum_k k n(k,t) = E_0 + m t
\end{eqnarray}
where the degree of each vertex is $k$ and the number of vertices
of degree $k$ at time $t$ is $n(k,t)$, the degree distribution.
The probability degree distribution is just $p(k,t) =
n(k,t)/N(t)$. The average degree $K$ tends to a constant with
\beq
 \lim_{t \ra \infty} K(t)
 = \lim_{t \ra \infty} \frac{2E(t)}{N(t)}
 =  \frac{2m}{\epsilon}
\eeq

The new edges added have one end attached to any new vertex if its
created, then the remaining ends are attached to vertices of the
existing graph chosen with the attachment probability $\Pi$. In
the mean field approach, we assume that the average value for the
degree distribution at any one time can be described by what happens
to the graph on average. This also means that all the parameters
$\epsilon, m$ could represent an average value for each time step,
and the equations are still an approximation to such a growth. The
evolution of the degree distribution is given in such a mean field
approximation by
\begin{eqnarray}
n(k,t+1) - n(k,t)
 &=& r
   [-n(k,t)\Pi(k,t) + n(k-1,t) \Pi(k-1,t) ]
   \nnel
   && + \epsilon \delta_{k,m} .
\label{neqn}
 \\
 r &:=& [(1-\epsilon)2 m + \epsilon m ], \label{rdef}
\end{eqnarray}

For the sake of simplicity, we will take the simple and often
studied form for the attachment probability $\Pi$
\beq
\Pi = p_v\frac{1}{N} + (1-p_v) \frac{k}{2E}
 \label{genprefatt}
\eeq
This represents a combination of random and preferential
attachment, such that existing vertices are chosen at
random\footnote{If we do not specify, then random means we draw
randomly from a uniform distribution.} $p_v$ of the time (first
term), while preferential attachment is used $(1-p_v)$ of the time
(second term).  Note that both terms require \emph{global}
information on the network through their normalisations.

The network evolution is therefore governed by four parameters,
$r,m,\epsilon$, and $p_v$. However, for almost all numerical runs
we will work with $\epsilon=1$, $p_v=0$ which corresponds to pure
preferential attachment in the mean field case.

With the attachment probability $\Pi$ of the simple form
\tref{genprefatt}, the mean-field equation can be solved exactly
in the long time, large $N$ limit.  It is also straightforward to
show that for a wider class of attachment
probabilities\footnote{Basically $\lim_{k\ra \infty} \Pi \propto
k$ is all that is required.} $\Pi$ the solutions tend towards a
power law form for large degree.  In particular for the form
\tref{genprefatt} one finds
\cite{KRL,KRR,WDN02,BAJ,DMS00,CNSW,KR}.
\bea
\lim_{k\ra \infty} \lim_{t \ra \infty} p(k,t) &\propto&
 k^{-\gamma},
  \label{psol}
 \\
 && \gamma = 1 + \frac{1}{p_v (1 - \frac{\epsilon}{2} )}
 \label{ggensol}
\eea
Since we study growing networks, $0 < \epsilon \leq
1$\tnote{$\epsilon =0$ is the pure rewiring case.}, and since now $0 < p_v
\leq 1$, we have that $2<\gamma<\infty$. The lower limit of the
power, $\gamma=2$, can be linked to the requirement that the
average degree is finite, that is the first moment of the
probability degree distribution  $K = [\int dk \, k p(k)]
/ [\int dk \, p(k) ]$ is finite. As $p_v \ra 0$ we get attachment to
vertices chosen randomly, and the distribution turns into an
exponential,
\beq
 \lim_{k\ra \infty} \lim_{t \ra \infty} p(k,t)
 \propto \exp\left\{ \frac{\epsilon}{r} k \right\}.
 \label{expsol}
\eeq
Although the attachment is random, this is not a standard
Erd\H{o}s-Reny\'i random graph.

Note that \tref{ggensol}  is a long time, large N solution.
However, all numerical models and all data sets are of finite size.
This introduces some natural scales and one would expect these to
lead to deviations from a simple power law in practical examples.
At low degree, the minimum number of edges added to a new vertex
(here $m$) sets such a scale. However, most power laws refer to
the large degree behaviour. There, for a real system, the
continuous part of the spectrum ends around $k_\mathrm{cont}$,
which can be defined through
\beq
p(k_\mathrm{cont}) = \frac{1}{N}
 \label{kcontdef}
\eeq
That is for $k \gtrsim k_\mathrm{cont}$ there will be some degree
values in any one example with no vertices of that degree.
Likewise, for $k \lesssim k_\mathrm{cont}$, we expect all
$n(k)>0$.  If we have a power law distribution, $k_\mathrm{cont}$
should scale as $k_\mathrm{cont} \propto N^{1/\gamma}$. Another
large scale exists for long tailed distributions, such as a power
law, where there are vertices with degree $k \gg k_\mathrm{cont}$.
For instance, the vertex of largest degree is the rank one vertex,
and its degree is likely to be $k_1$, where
\beq
 \sum_{k=k_1}^\infty p(k) = \frac{1}{N}
 \label{k1def}
\eeq
This scales as $k_1 \propto N^{1/(\gamma-1)}$ for a power law
distribution.

An approximate analytic finite time or size solution to the mean
field equation \tref{neqn} for the case of pure preferential
attachment with number of edges equally the number of vertices
(here $m=1$, $p_v=0$, $\epsilon=1$) was given by Krapivsky and
Redner \cite{KR02} (see also \cite{DMS01,KK00,ZM00}). The form is
\bea
p(k,t) &=& p_\infty(k) F_s(k,t)
 \label{ddmfsol}
\\
p_\infty(k) &=& \frac{2m(m+1)}{k(k+1)(k+2)}
 \label{ddmfinfsol}
\eea
Asymptotically the finite size scaling function $F_s$ is a
function of $x=k/(2t^{1/2})$ and it differs from one only for $x
\lesssim 1$. With $\gamma=3$ for this case, we have that $N \sim t
\sim (k_1)^2$ so $F_s \not\approx 1$ only for $k \gtrsim k_1$.  It
also follows that it is sensitive to initial conditions since the
vertices of biggest degree are the oldest.  For the initial
conditions $n(k=m,t=1)=2$ $n(k\neq m,t=1)=0$ and  generalising the
arbitrary $m$ but keeping pure preferential attachment ($p_v=0$,
$\epsilon=1$), we use the approach of \cite{KR02} to find that
\bea
F_s(k,t) & \approx &
 \mathrm{erfc}(x) + \frac{e^{-x^2}}{\sqrt{\pi}}
 \left(
 2x + \sum_{n=3}^{m+2} \frac{8}{n!} \left( 1 +
 (1+m)\delta_{m+1,n} \right) x^n \: \mathrm{H}_{n-3}(x) \right)
 \label{Fsdef}
\eea
and it is made up of the complementary error function
$\mathrm{erfc}$ and Hermite polynomials $\mathrm{H}_n$.

\begin{figure}[!htb]
\begin{center}
 \scalebox{0.6}{\includegraphics{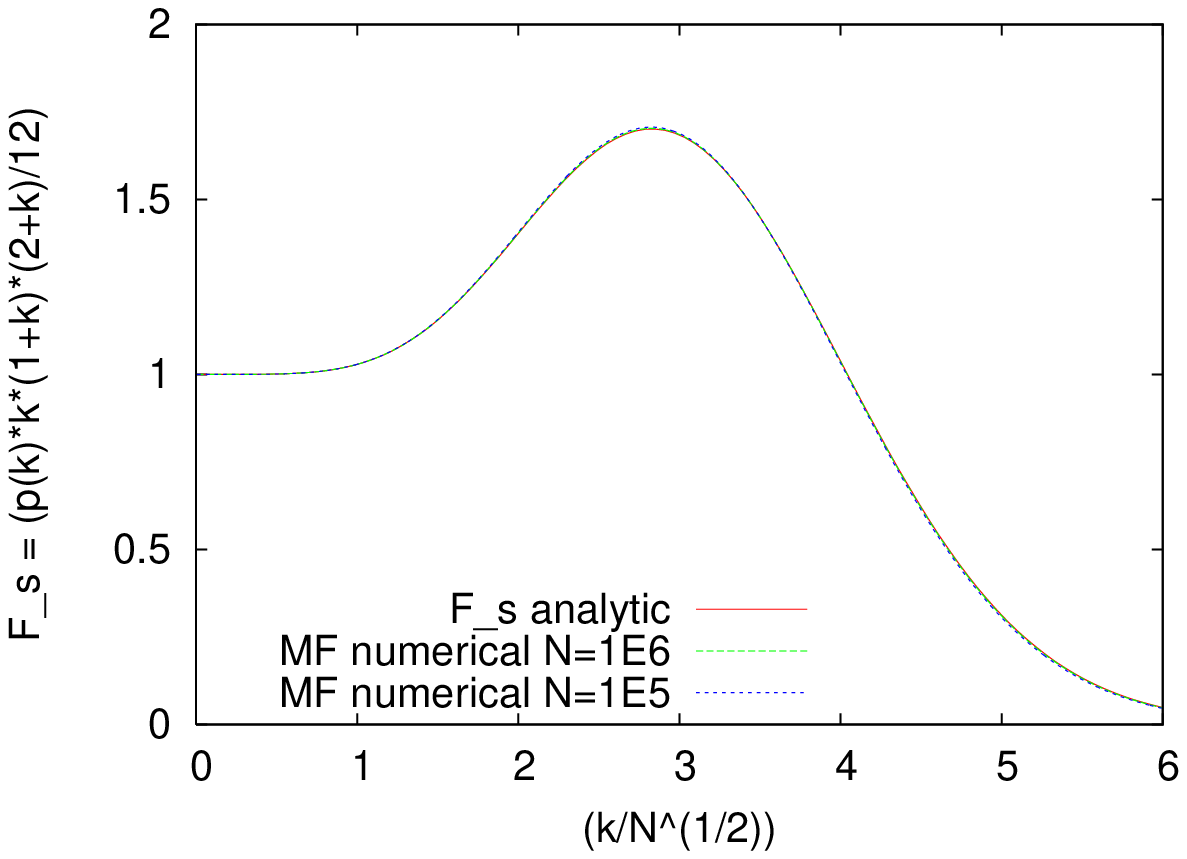}}
 \scalebox{0.6}{\includegraphics{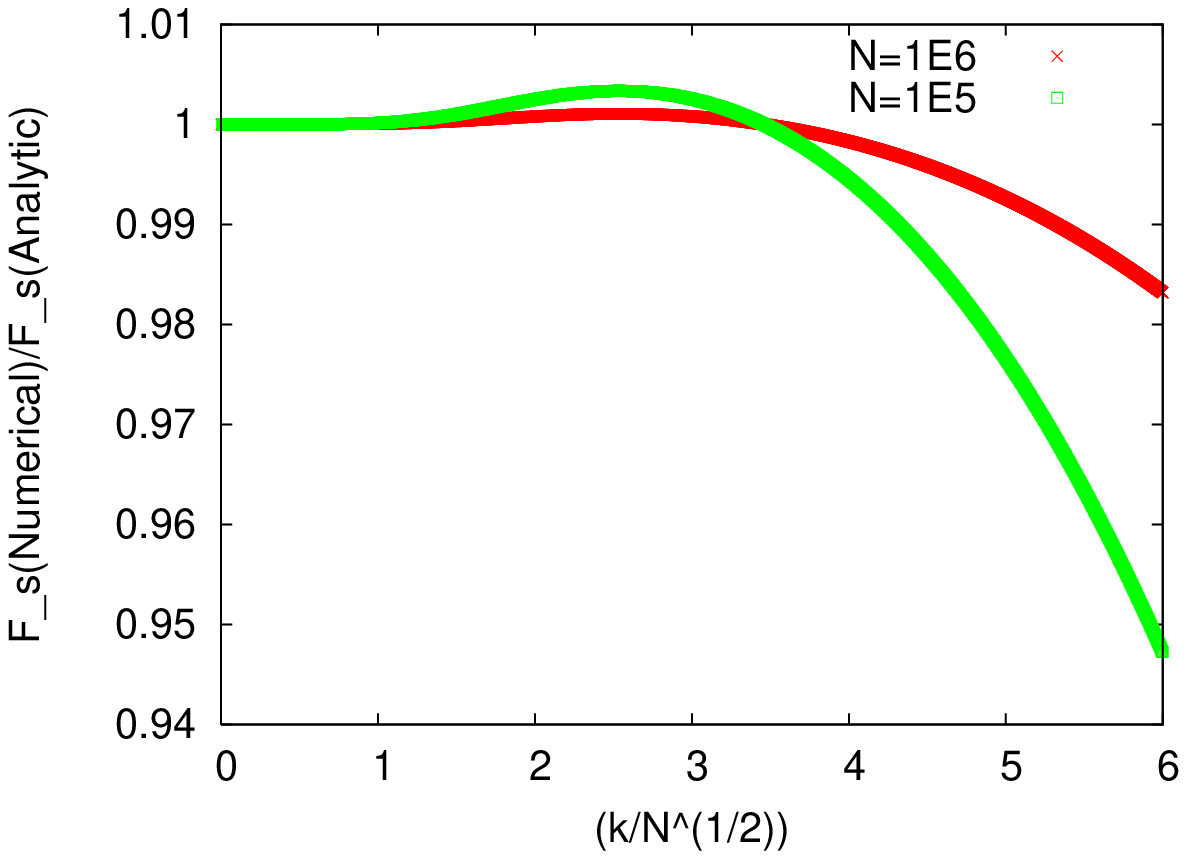}}
 \end{center}
\caption{On the left the mean field results, analytic solution,
and numerical solutions for $N=10^5$ and $N=10^6$, plotted to show
the form of the scaling function, all for $m=2$, $\epsilon=1$,
$p_v=0$.  No difference is visible in this plot so on the right
the numerical data is plotted divided by the analytic solution.}
 \label{fmfsol}
\end{figure}

The analytic form of the finite size function $F_s$ \tref{Fsdef}
is a good approximation to that found from a direct numerical
solution of the mean field equations as figure \ref{fmfsol} shows.

\section{The Generalised Walk Algorithm}\label{swalk}

The mean field equations \tref{neqn} can be implemented in a
straight-forward manner, by choosing vertices in the existing
graph at random using the probability $\Pi(k)$ implemented
explicitly in an algorithm.  This is done in most cases.  As
discussed in the introduction, the walk algorithm provides a
natural mechanism for such a probability to emerge naturally from
an intrinsic property of the graph.  The basic walk algorithm we
will consider is merely a generalisation of the original
Saram\"aki and Kaski \cite{SK04} algorithm\footnote{Preliminary
studies of such models were also made independently by one of us,
TSE, in collaboration with Klauke \cite{Kla}.}:
\begin{enumerate}
\item Start with any graph\footnote{In fact, the way the algorithm is
phrased we require that no vertex has zero degree but with a small
adjustment even this limitation could be dropped.} $G(t=0)$ and
start the time counter at $t=0$.
\item With probability $\epsilon$ choose to
add a new vertex $v_0$. The remaining time, let $v_0$ be a random
vertex in the graph chosen with probability\tnote{Note that in the
mean field equations we have only one probability for choosing
vertices in the existing graph at whatever point in the algorithm
we are considering.} $\Pi$. Now start adding new edges, counting
from $i=1$.
\item To start the random walk we choose a vertex $v_i$ in the existing graph,
$G(t)$.  We will consider several different ways to do this.
\item Now make one step in a random walk on the graph by choosing
one of the neighbours of $v_i$ at random\footnote{One can vary
this aspect.  By using a biassed walk, say choosing neighbours
preferentially based on colour of vertices or weights of edges, or
based on other vertex properties such as the degree or clustering
of the target, one might get interesting variations.}. Move to
this neighbour and now set $v_i$ to be this vertex.
\item Repeat the previous step $l$ times.
\item Repeat from step three $m$ times, increasing $i$
each time $i=1,2,\ldots,m$.
\item Now create $G(t+1)$ by adding vertex $v_0$ and the edges $\{ (v_0,v_i)
| i = 1,2,\ldots,m\}$ to the graph $G(t)$.  At this point one
might also choose to reject some of potential edges and maintain
some characteristic of the graph.
\item Increase $t$ by one and repeat from second step.
\end{enumerate}

There are several variations within the general algorithm which we
will study.  We will indicate our choices by the binary bits of a
parameter $\texttt{v}$.
\begin{enumerate}
\item[A] The walks can be started from a vertex chosen
randomly ($\mathtt{(v \& 1) =1}$), as done in \cite{SK04}, or by
taking a random end of a random edge ($\mathtt{(v \& 1) =0}$).
\item[B] One could start a new walk for every new edge
($\mathtt{(v \& 2) =1}$).
Alternatively, as in \cite{SK04}, we could start a new walk at
each time step, the $i=1$ edge,  but then we take the end of the
previous walk $v_{i-1}$ to start the walk for the $i$-th edge
($\mathtt{(v \& 2) =0}$).
\item[C] The length of the random walks can be fixed to be $l$ as
in \cite{SK04}, ($\mathtt{(v \& 4) =0}$). This might not be
realistic in many cases so we have also looked at the case where a
further step on the walk is made with probability $p_l = l/(1+l)$
so that the average walk length was $l$ ($\mathtt{(v \& 4) =1}$).
\item[D] The number of edges could be fixed to be $m$ at each time step as in
\cite{SK04} ($\mathtt{(v \& 8) =0}$). This could be varied in a
similar manner to the walk length, with one edge always added (to
ensure a connected graph) but subsequently another edge is added
with probability $p_e=(m-1)/m$ so on average $m$ will be added
($\mathtt{(v \& 8) =1}$).
\end{enumerate}

Intuitively, the initial point of the random walk should be
immaterial for `long' walks.  In \cite{SK04} it was indicated that
for their algorithm (essentially the $\mathtt{(v \& 1) =1}$ choice
here) long was just one step\footnote{The General Network with
Redirection model in \cite{KR,KR02} is similar to our single step
walks with a stochastic element $\mathrm{(v\& 4)} =1$, and there
good power laws were also noted.}. Presumably, this indicates that
there is already little correlation between the connectivity of
nearest neighbour vertices, and it is this correlation length,
rather than mean shortest separation or diameter length scales,
which is important. This is also an assumption behind the
mean-field approximation, so we should expect that the mean field
equations are a good approximation to graphs produced from random
walk algorithms. This will be confirmed below.

For the stochastic choices in options C and D, the Markov process
used here produces a large peak at small values.  Thus for the
walks of random length in case C, a fraction $(1-p_l)$ vertices
are attached to the vertex at the start of the walk.  If this
initial vertex is chosen randomly ($\mathtt{(v \& 1)} =1$ in
option A), and given that one step is often sufficient to produce
reasonable scale-free behaviour, then we are actually reproducing
the mixed preferential attachment and random attachment algorithms
mentioned above with $p_v \sim (1-p_l) = 1/(1+l)$. This is yet
another way that a walk algorithm might produce various powers
$\gamma$ as \tref{ggensol} indicates.  Many other distributions
could be tried for stochastic choices so the Markov process used
here is merely exemplary.

If the length of the walk is zero then we get some special
behaviour. If we choose the vertices $v_i$ at random, we are then
generating a graph with an exponential distribution for $n(k)$
\tref{expsol}. On the other hand,
choosing to connect to vertices in the existing graph
by choosing the random end of a random edge is guaranteed to
generate a scale-free graph as noted in \cite{DMS01}. Thus we
expect that with this start for the random walks, all graphs are
scale-free whatever the walk length. \tnote{Another variation would be
to add a second choice to each step of the walk.  That is we
either follow a random edge or, with probability $p_v$ we choose a
random vertex, essentially restarting the walk.  Then this is
essentially the $p_v$ of the mixed preferential attachment and
random attachment \tref{genprefatt} and so is another way to vary
the power in scale-free graphs produced from random walk
algorithms.  This we do not pursue here.}

Finally, we note that one might often wish to limit the graphs
generated to be simple, with no multiple edges between vertex
pairs and no edges with the same vertex at both ends. We have done
numerical simulations both with and without this limitation, and
found that for $N=10^6$ and other typical values used here, the
difference is negligible with a very small fraction of edges
rejected\footnote{In one run with an implementation of an
algorithm exactly as stated, so allowing multiple edges and edges
connected to one vertex only, with $N=10^6$ vertices and $E=2
\times 10^6$ edges, using a walk of fixed length of 7 steps and
starting a new walk from a random vertex for every new edge added,
and $\epsilon=1$, there were just 76 double edges produced, with
no triples or higher. In \cite{SK04} the graph generated was
simple.}.

\section{Results for Unweighted Graphs}

\subsection{Degree Distributions}

\begin{figure}[t]
\begin{center}
 \scalebox{0.6}{\includegraphics{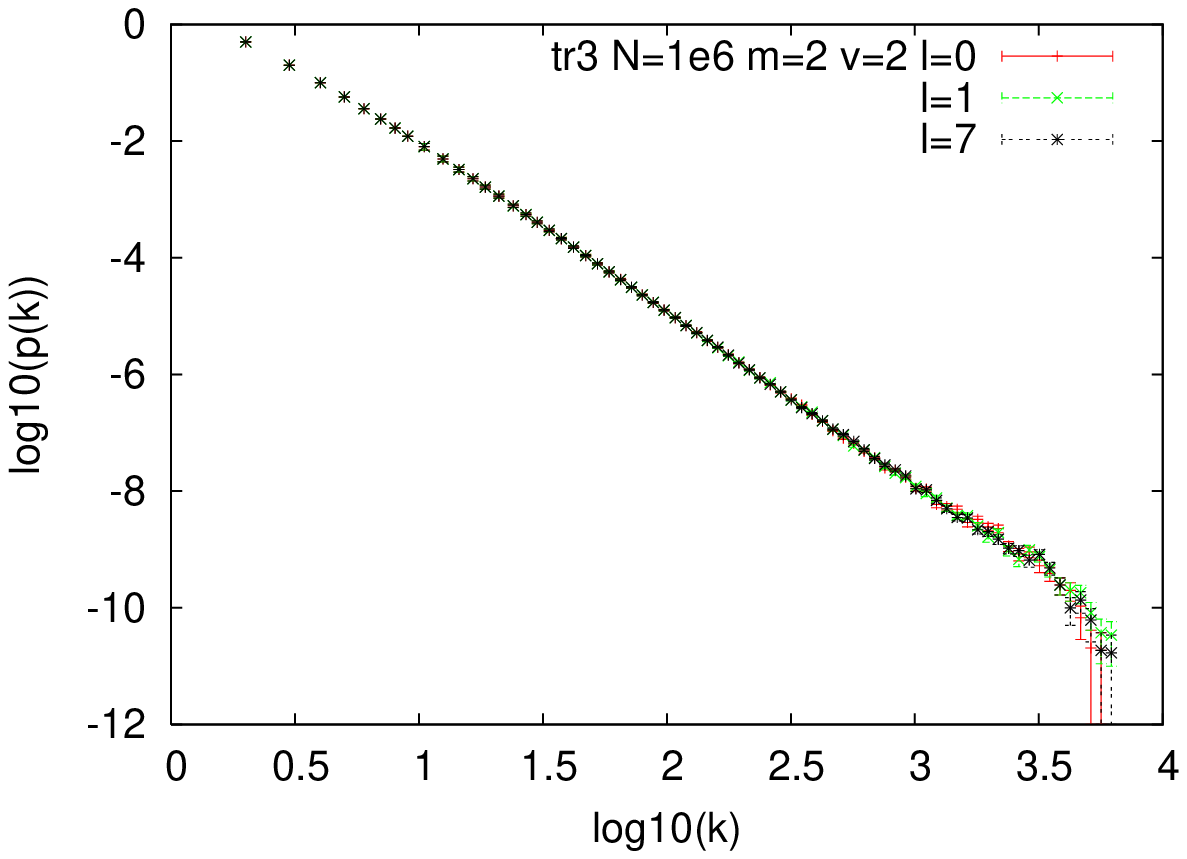}}
 \scalebox{0.6}{\includegraphics{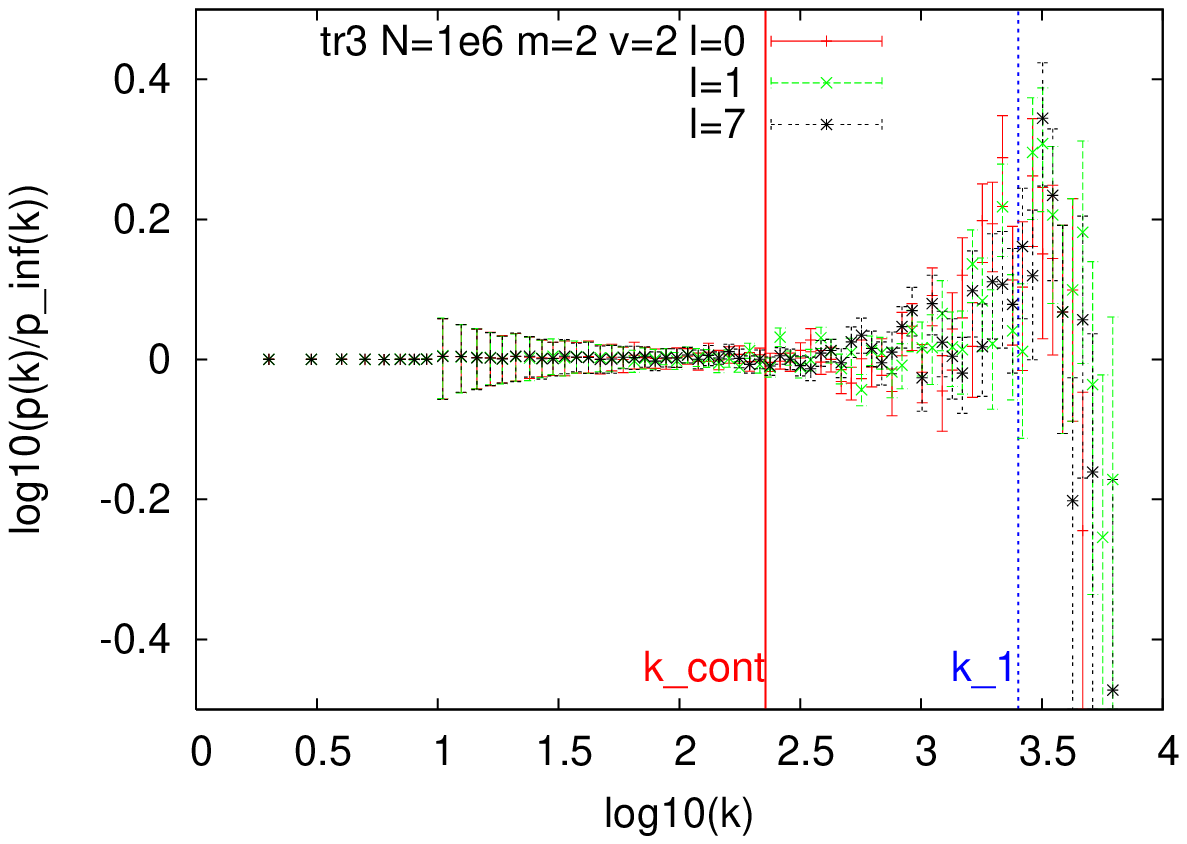}}
\end{center}
\caption{Degree distributions for networks of size $N=10^6$,
generated using random walks started from a random end of a
randomly chosen edge. The left panel displays the raw degree
distribution, and the right the degree distribution normalised by
equivalent mean field $t\rightarrow\infty$ solution $p_\infty(k)$,
with finite size correction $F_s$ visible for $k>k_\mathrm{cont}$.
All variations with this initial condition  ($\mathtt{v \& 1} =0$)
show the same behaviour. Here,  one vertex ($\epsilon=1$) with two
edges ($m=2$) are added per time step. The results are shown for
average walk lengths of 0,1 and 7 steps, with data averaged over
100 runs.
In this example, a new walk is started for every new edge added.  }
 \label{ftr3n1e6s017v2}
\end{figure}

First, we will note how robust the walk algorithm is at producing
scale-free networks.  Figure \ref{ftr3n1e6s017v2} shows the degree
distributions for an exemplary walk algorithm which started all
random walks from a random end of a randomly chosen
edge.\tnote{The results of \tref{ftr3n1e6s017v2} are from version
3 of the \texttt{JAVA} programme \texttt{timwalk}.} This
is equivalent to pure preferential attachment if no walk is
made ($l=0$). Longer walks or other variations in the algorithm
do not alter this result.

\begin{figure}[!tb]
\begin{center}
 \scalebox{0.6}{\includegraphics{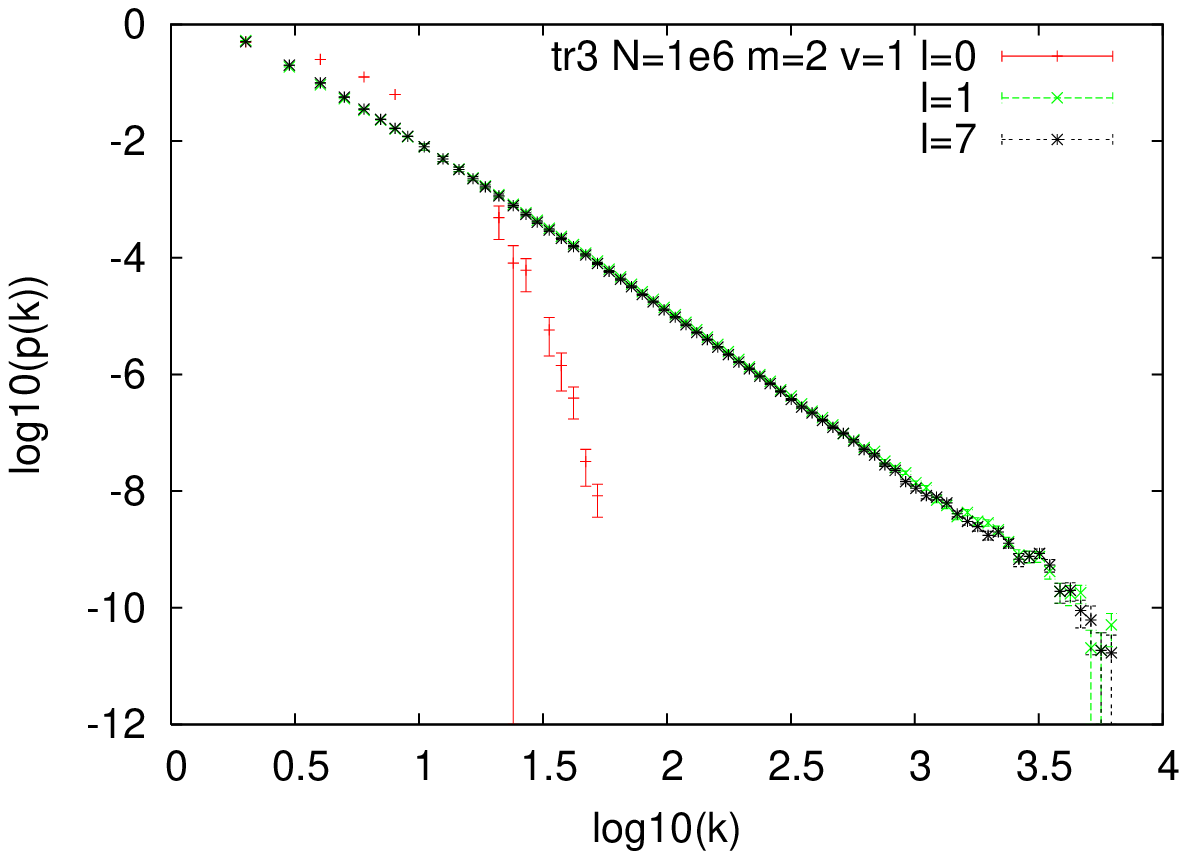}}
 \scalebox{0.6}{\includegraphics{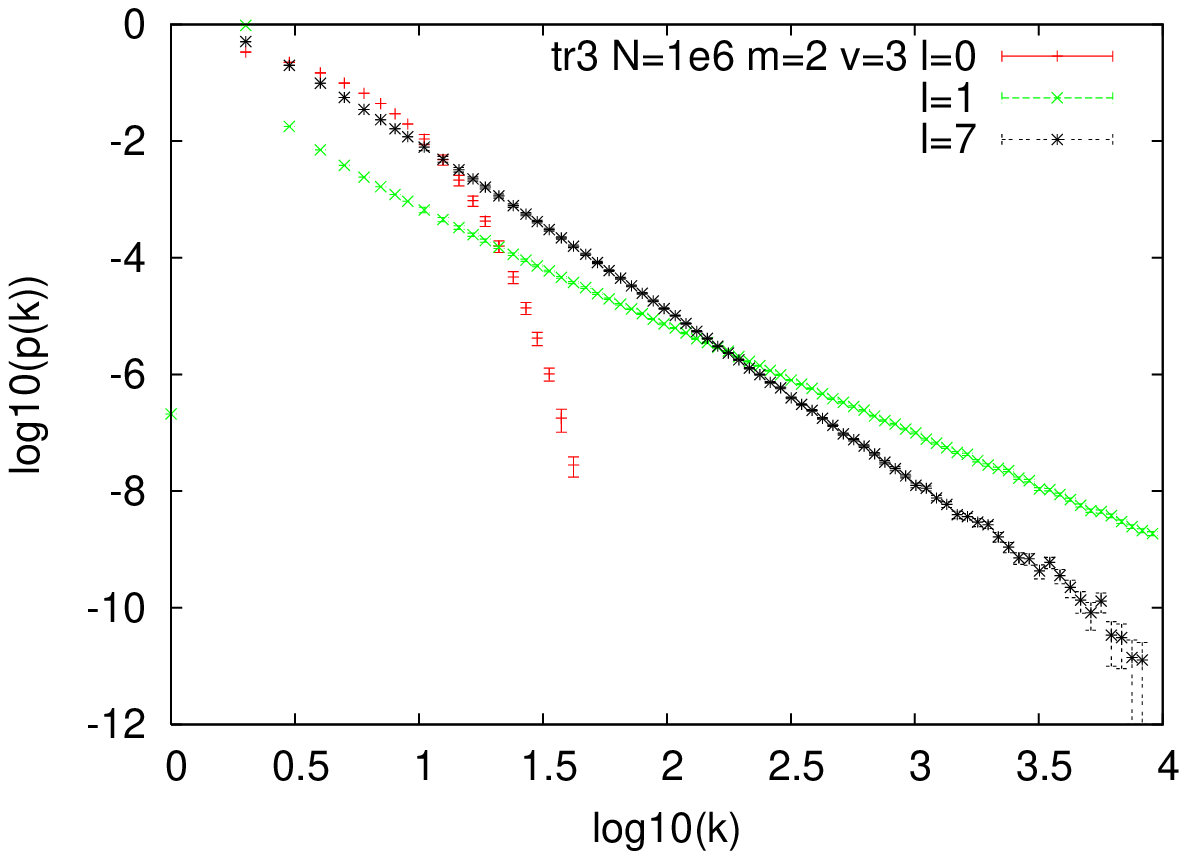}}
\\
 \scalebox{0.6}{\includegraphics{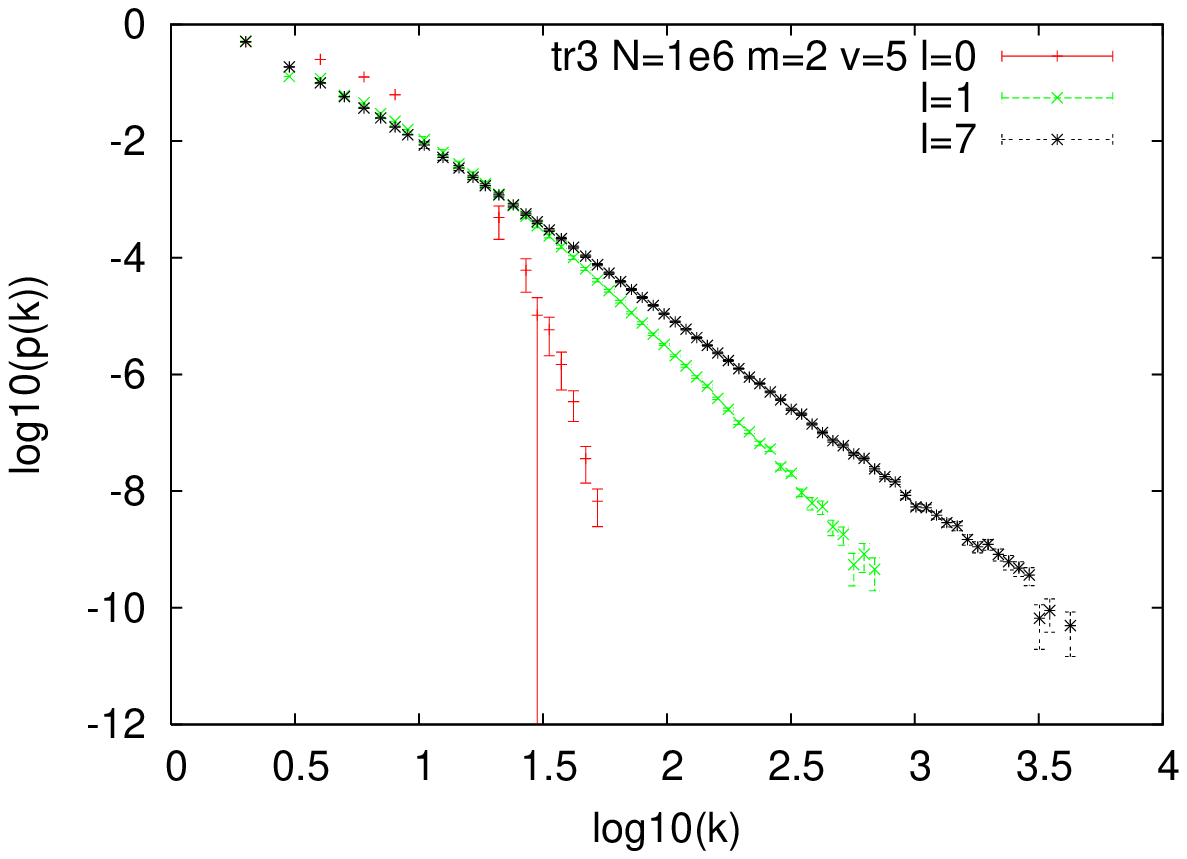}}
 \scalebox{0.6}{\includegraphics{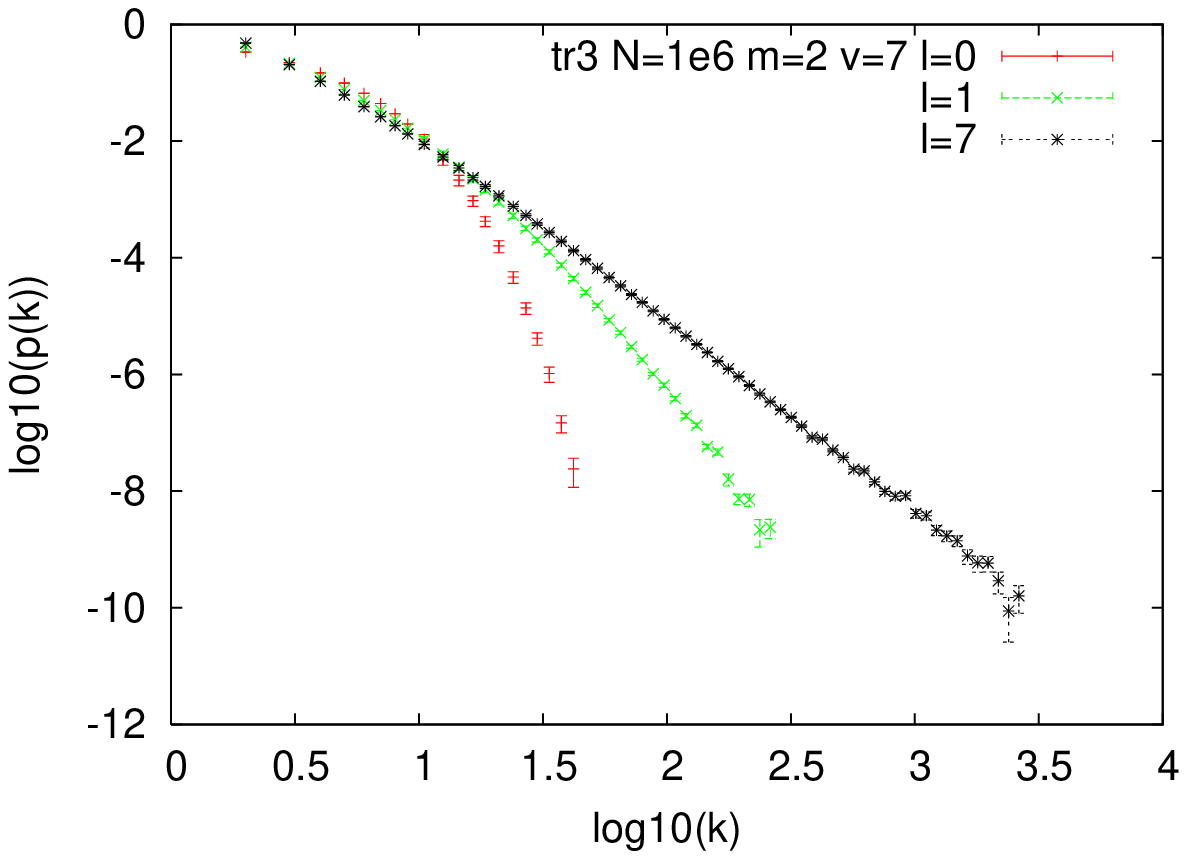}}
\\
 \scalebox{0.6}{\includegraphics{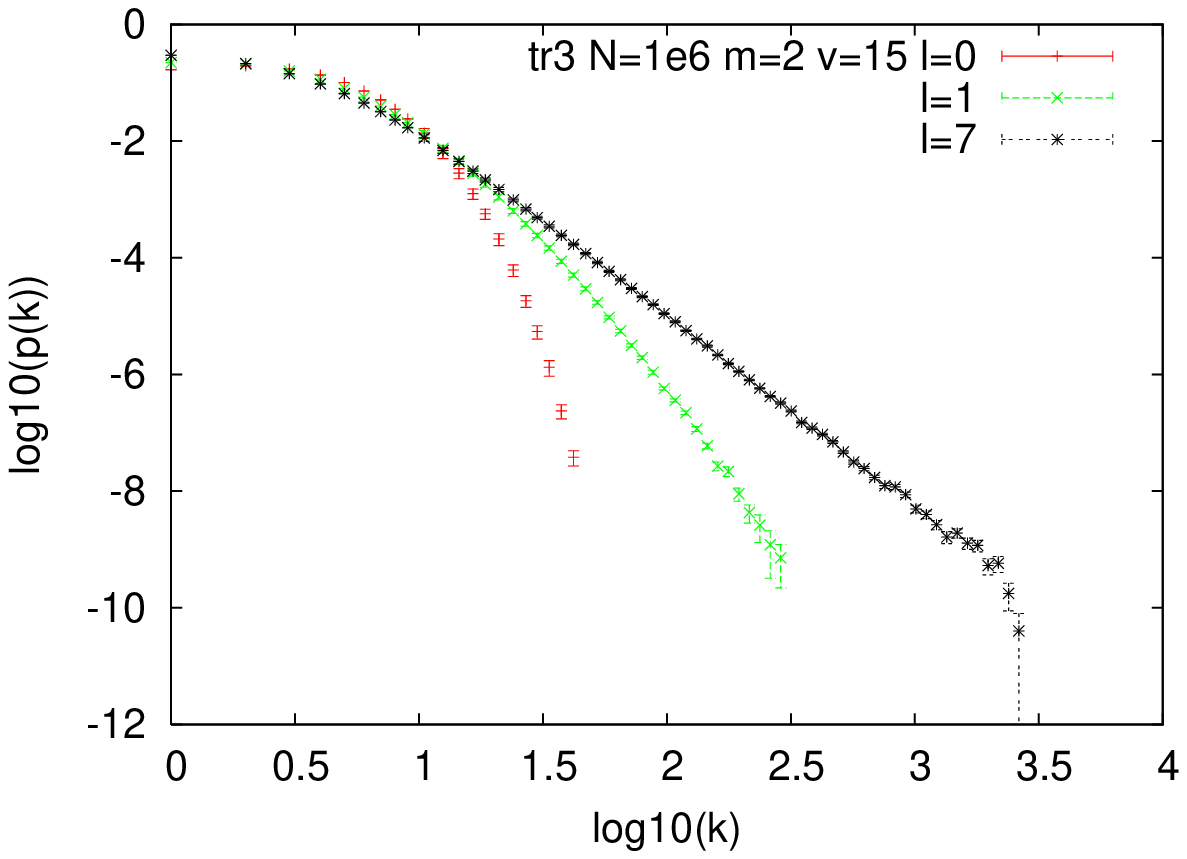}}
 \end{center}
\caption{Plots of $\log_\mathrm{10}(n(k))$ vs
$\log_{\mathrm{10}}(k)$ for $N=10^6$ networks generated by random
walks started from a randomly chosen vertex  ($\mathtt{v \& 1} =1$),
with one vertex ($\epsilon=1$) and two edges ($m=2$) added at each time step.
In each graph, the results are shown for average walk lengths $l$ of 0, 1 and 7 steps,
with data averaged over 100 runs. In the top row, the walk length $l$ is fixed,
whereas in the middle row the length is chosen using a Markov process.
In the left column all $m$ new edges are attached to
vertices chosen by one continuous walk, whereas in the right column
a new walk is started for each edge added. The bottom figure has variable numbers
of edges and variable walk length. Multiple edges are allowed here.}
 \label{ftr3n1e6s017v1357}
\end{figure}

\begin{figure}[tb]
\begin{center}
 \scalebox{0.8}{\includegraphics{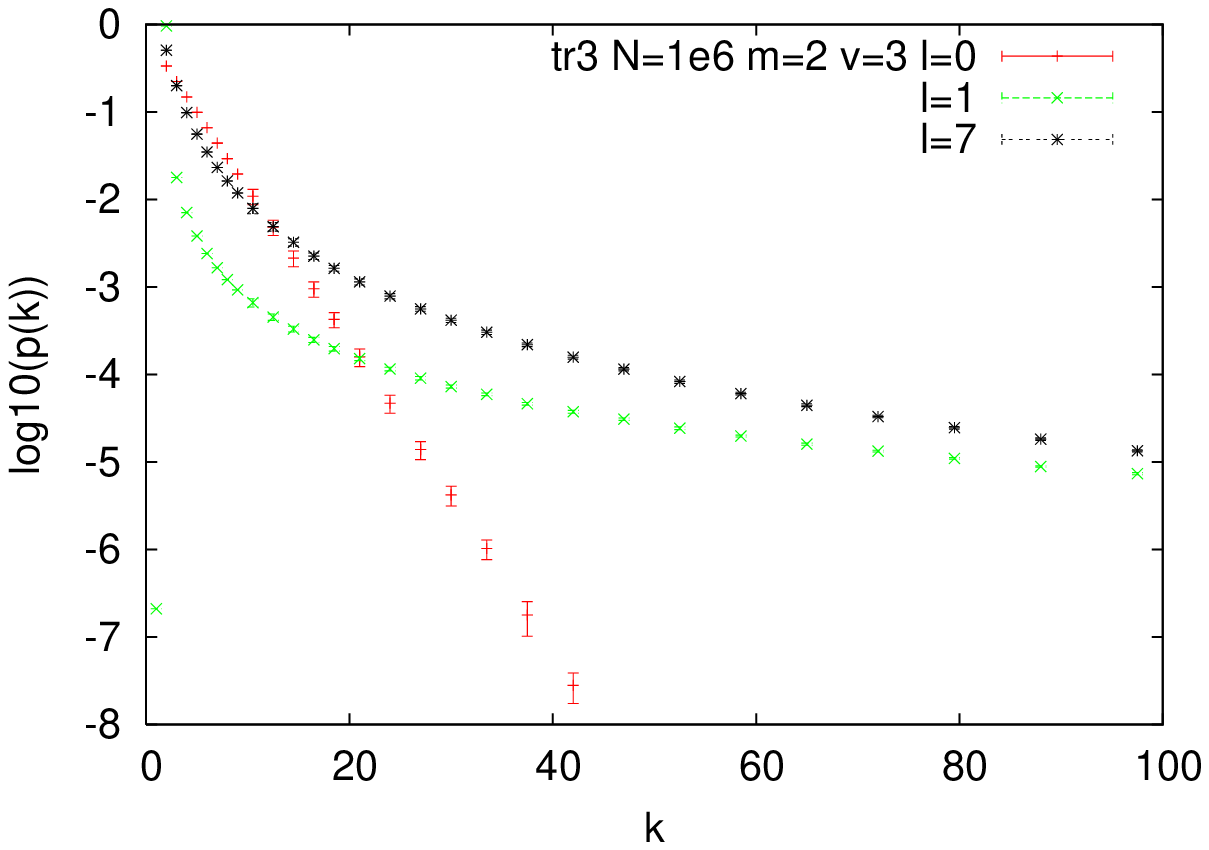}}
 \end{center}
\caption{Plot of $\log_\mathrm{10}(n(k))$ vs $k$ for $N=10^6$ networks,
generated using walks of
fixed length started from a randomly chosen vertex for each new
edge ($\mathtt{(v \& 1) =3}$), with $\epsilon=1$ and $m=2$, and
$l=0,1,7$. Data are averaged over 100 runs. Multiple edges are allowed here.}
 \label{tr3e1000000s017k2v3r_logbin_av_sl}
\end{figure}

More revealing are algorithms which start their walks from a
randomly chosen vertex as seen in figure \ref{ftr3n1e6s017v1357}.
As expected from the mean field approximation, starting from a
random vertex but doing no walk ($l=0$) produces an exponential
distribution seen by the very short tailed distribution in all
cases for the $l=0$ lines of figure \ref{ftr3n1e6s017v1357}.  This
is also illustrated in the semi-log plot of Fig.~\ref{tr3e1000000s017k2v3r_logbin_av_sl}.
On the other hand, any walk of $l\geq 1$ produces a distribution
with a power-law-like tail that is much longer than the
exponential distributions \tref{expsol} of the zero step walks.
The ($v=1$) variant of the algorithm, where a new walk is started
only for every new vertex, with $l$, $m$, and $\epsilon$ fixed,
produces very consistent degree distributions for $l\geq 1$
(Fig.~\ref{ftr3n1e6s017v1357}, top left panel). This is
essentially the algorithm used by Saram\"aki and Kaski
\cite{SK04}. When $l$ is small, other variations of the walk have
an effect on the slope of the degree distribution. In particular,
the variants using a Markov process for a single step walk (e.g.\
$l=1$, $v=15$) fit a power-law in their tails which is closer to
$\gamma=5$ (Fig.~\ref{ftr3n1e6s017v1357}, bottom panel). This
value corresponds to the earlier discussion, where a probability
$(1-p_l)$ of making a zero step walk from a random vertex start
(in option C) can be taken as a first approximation to be
equivalent to the probability $p_v$ for random vertex attachment
in the mean field equations \tref{genprefatt}.\tnote{As here half
of the time on an $l=1$ walk we will be attaching to the randomly
chosen vertex at the start of the walk. The power expected from
the large $N$ limit is in this case $\gamma =5$ from
\tref{ggensol}.} Our one step Markov walk results (cases $l=1$ and
$v=5,7,15$ in figure \ref{ftr3n1e6s017v1357}) support this and
will be considered again with figure
\ref{tr3e1000000s1k2v135715gamma} below.  Likewise the variation
with the length of walk $l$ is also shown in figure
\ref{ftr3scaling} below and different algorithms for the same long
seven step walks, figure \ref{gammam2l7v135715} will be discussed
in more detail below.

In the case of $l=1$, starting a new walk from a randomly chosen vertex for
each of the $m$ new links ($v=3$) (Fig.~\ref{ftr3n1e6s017v1357}, top
right panel) appears to result in a
much smaller power than $\gamma=3$,
unlike in the ($v=1$) case where the vertices are selected
using one continuous walk. This is possibly because in
the $v=3$ algorithm all vertices chosen are only one step away
from a randomly chosen vertex, while in the $v=1$ case \cite{SK04},
one vertex is one step and the other two steps, on average 1.5 steps,
from a randomly
chosen vertex.  This suggests that there are weak correlations
between properties of neighbouring vertices, but not between next
to nearest neighbours.  Thus the effective longer range of a $v=1$
one step walk over a $v=3$ one step walk accounts for the
differences between these two variants.

Certainly, the longer the walk, the more the distributions become
identical, whatever the details of the algorithm for our large
$N=10^6$ networks, with tails approaching a power law with powers
around $\gamma=3$.

\begin{figure}[!htb]
\begin{center}
 \scalebox{0.8}{\includegraphics{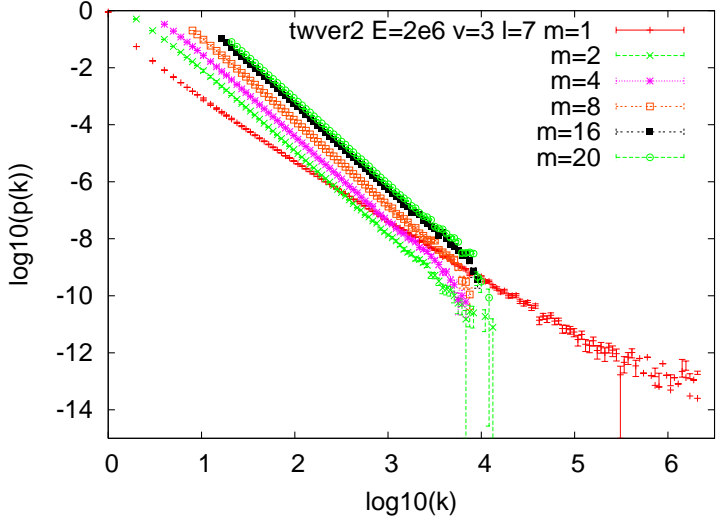}}
 \end{center}
\caption{The normalised degree distributions,
$\log_\mathrm{10}(p(k))$ vs $\log_{\mathrm{10}}(k)$, for fixed
number of edges $E=2\times 10^6$,  $\epsilon=1$, and varying
average degree $m$. For random walks starting from a random vertex
for every new edge and of fixed length $l=7$. Averaged over 100
runs.}
 \label{ftwver2E2e6}
\end{figure}

Varying the average degree $K$, but holding the number of edges
fixed shows nothing of note except when $m=1$, i.e.\ where we
generate a tree graph with no loops, as one can see in figure
\ref{ftwver2E2e6}.\tnote{The results of \tref{ftwver2E2e6} are
from version 2 of the \texttt{JAVA} programme \texttt{timwalk}.}

\subsection{Finite-Size Effects}

\begin{figure}[!htb]
\begin{center}
 \scalebox{0.8}{\includegraphics{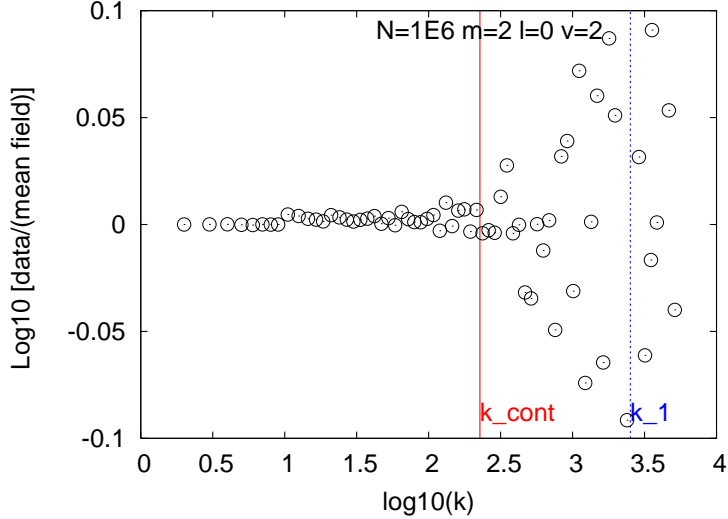}}
 \end{center}
\caption{Degree distribution from random walk
algorithm ($N=10^6$,  $\epsilon=1$, v=2, l=0, m=2,
averaged over 100 runs) normalised by the
numerical solutions to the mean field equations.
The vertical lines indicate the characteristic
scales $k_\mathrm{cont}$ (left) and $k_1$ (right).}
 \label{fmfdatafit}
\end{figure}

The degree distributions discussed above are not simple power
laws. This is to be expected since the solutions to the mean field
equations do not predict this as \tref{ddmfsol} shows. Also the
mean field equation is itself an approximation, but it should be
closest to models with genuine preferential attachment.
Fig.~\ref{fmfdatafit} displays the degree distribution for
networks generated with algorithms where the random walks start
from an end of a randomly chosen edge ($\mathrm{(v \& 1)}=0$),
compared against the numerical mean field solutions. The data fits
the finite $N$ mean field solutions well, with the deviation from
mean field comparable to the apparent statistical variation and
systematic effects from the logarithmic binning. However, its
clear that the data has large fluctuations and so is poor for
large degree, $k > k_\mathrm{cont}$.\tnote{Show for $v=1$ or $v=3$
or $v=15$ too. Also try to fit powers to low $s$ Markov process
examples.}\tnote{NOTE TO TIM: THIS IS AS FAR AS I GOT, HAVEN'T
TOUCHED ANYTHING THAT FOLLOWS! -Jari}

Given that the mean field solutions \tref{ddmfsol} are an
excellent representation of genuine preferential attachment
models, it is interesting to see if this is useful for the results
of all random walk models.
However, before we look at more data we need to consider the sizes
of the scales in our finite sized examples to understand
deviations from a pure power law. For large scales, $k \gtrsim
k_1$, modifications to a pure power law result from a finite size
correction similar to the $F_s$ \tref{Fsdef} found for pure
preferential attachment models. However, this correction is not of
practical importance by definition there is essentially no data
for $k \gtrsim k_1$. The data is best for  $k \lesssim
k_\mathrm{cont}$ of \tref{kcontdef} . In practice this scale is
not large, for a million vertex graphs (few data sets have bigger
graphs) $k_\mathrm{cont}$ is only\footnote{For the mean field
model solution \tref{ddmfinfsol} with $m=2$ the large scales are:
$k_\mathrm{cont}= 105$ and $k_1= 796$ ($N=10^5$),
$k_\mathrm{cont}= 227$ and $k_1= 2520$ and $N=10^6$.  In fact the
degree with local power $\gamma_\mathrm{eff}$ \tref{geffdef}
closest to the theoretical value is found just above
$k_\mathrm{cont}$ at $k_\mathrm{max}=149$ for $N=10^5$ while for
$N=10^6$ this is at $k_\mathrm{max}=388$.} of order 100. Thus most
data sets, and certainly our model runs, are actually mesoscopic
systems. It also means that there are significant deviations from
a power law because of the \emph{small scale} effects. For
instance the mean field large time solution \tref{ddmfinfsol}
shows deviations from the inverse cubic large degree behaviour for
degree scales $k \sim O(1)$. These small scale deviations are
finite $N$ effects in the sense that $k_\mathrm{cont}$ is finite
only for finite $N$ and is in practice close to one.

We can illustrate the problem by studying the mean field
solutions, fitting a power law to neighbouring points and
estimating the power $\gamma$ through
\beq
\gamma_\mathrm{eff} (k)
 = - \frac{\ln[p(k+1))/p(k)]}{\ln[(k+1)/k]} .
 \label{geffdef}
\eeq
In fact for pure preferential attachment models this effective
measure of the power law coefficient $\gamma$ is always below the
large $N$ value for any useful degree $k$ since using
\tref{ddmfinfsol} we have
\beq
 \gamma_\mathrm{eff} (k)
 = 3 \left( 1 - \frac{1}{k} + O\left(\frac{1}{k^2}\right) \right)
 \qquad (1 \ll k \ll k_1)
 \label{geffsmallk}
\eeq
For $N=10^6$ (larger than most data sets) $k_\mathrm{cont} \sim
100$ is the largest degree with useful data so we'd expect the
local power to be at least of order one percent below the large
$N$ value associated with the formation mechanism for the graph.
So even in this perfect pure preferential attachment model, simple
power law fits to reasonable data sets are going to underestimate
the power which in turn would lead to a misunderstanding of the
underlying formation mechanism, e.g.\ though formulae such as
\tref{ggensol}. In practice, results are likely to be worse than
this.

The discussion above highlights the problems in interpreting any
power fitted to finite $N$ data.  With these warnings in mind let
us now turn to more general random walk models and look at the
power law behaviour, focusing more on the comparison between the
various random walk algorithms.  We will also compare against the
appropriate numerical mean field equation solutions, for which we
have a complete understanding of the finite size effects.

\begin{figure}[!htb]
\begin{center}
 \scalebox{0.6}{\includegraphics{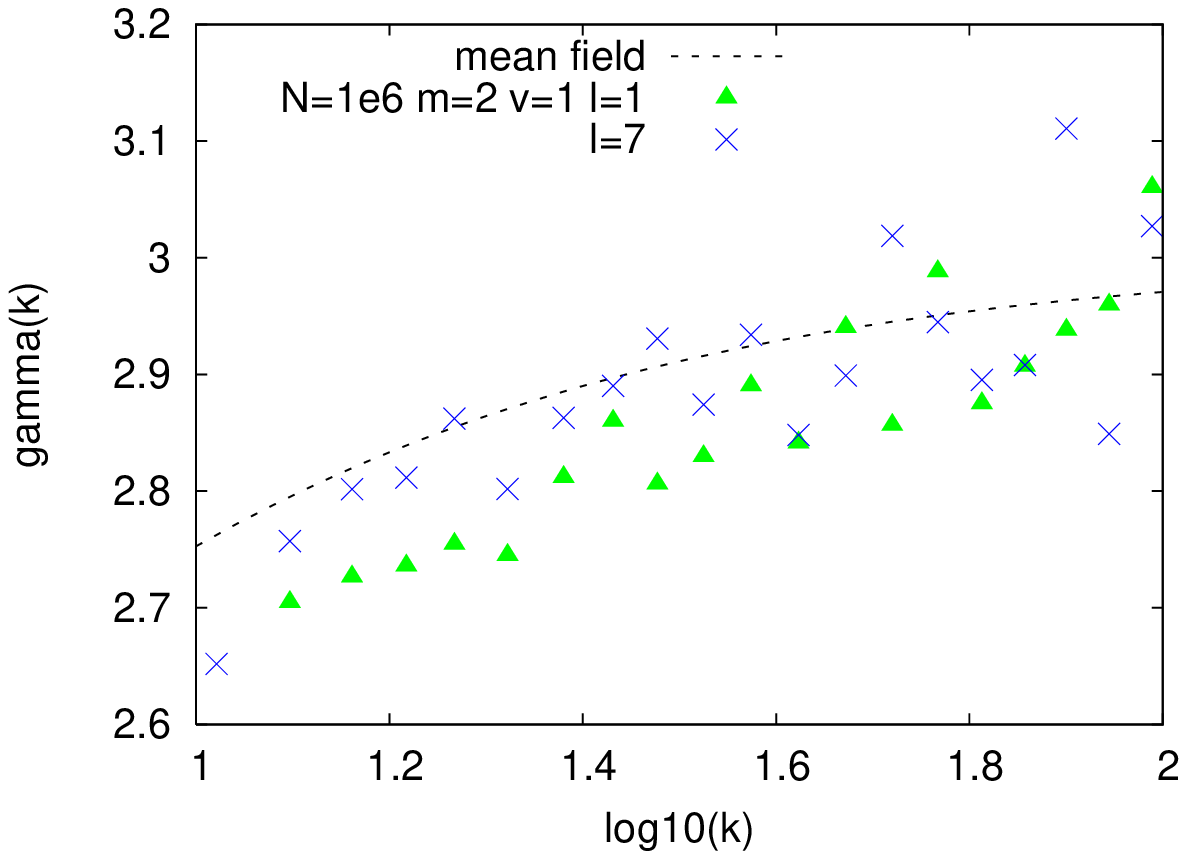}}
 \scalebox{0.6}{\includegraphics{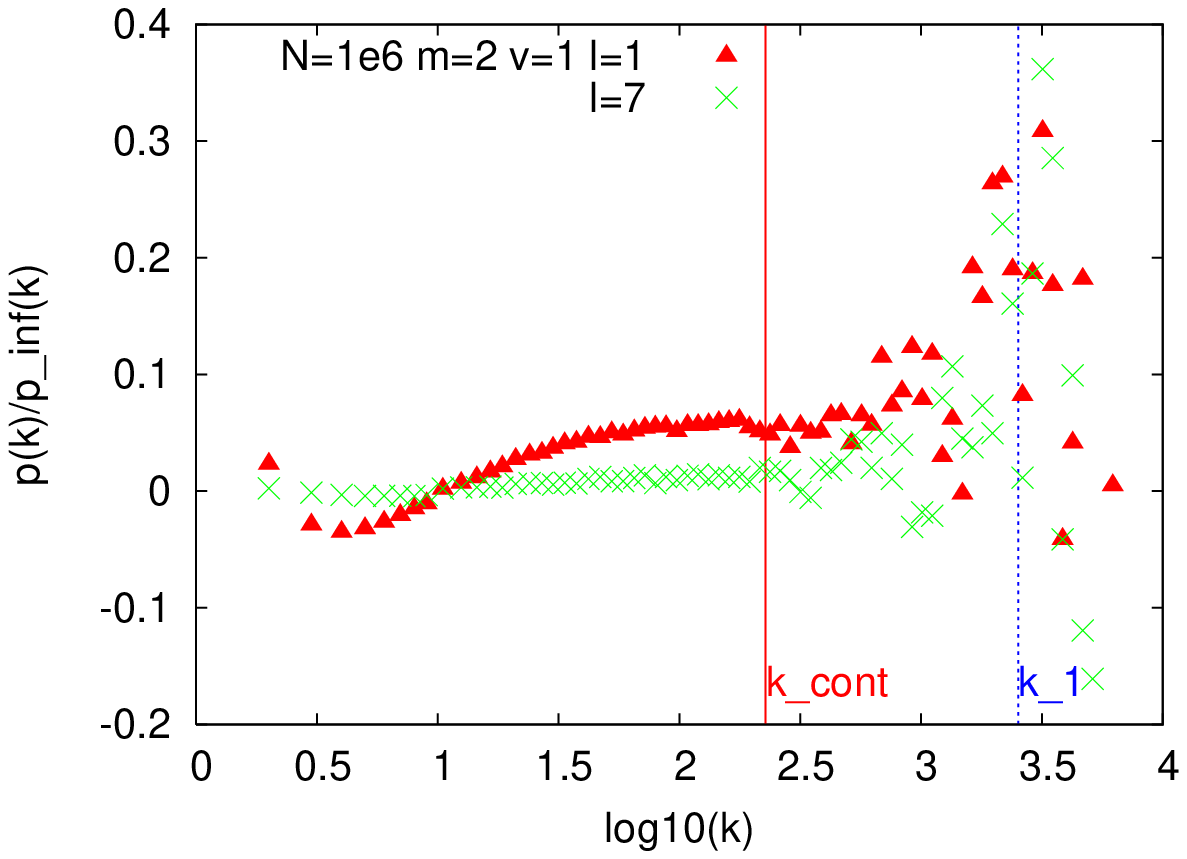}}
 \end{center}
\caption{Comparison of one and seven step walks for
Saram\"aki-Kaski style algorithm $N=10^6$, $\epsilon=1$, $m=2$
$v=1$.  The effective power $\gamma(k)$ on the left compared
against numerical mean field solution. On the right data is
normalised by the large $N$ mean field solution for graph of
similar characteristics.}
 \label{tr3e1000000s1k2v1gamma}
\end{figure}

First it is interesting to note that, while even short walks have
long tailed distributions that are well approximated by a power
law (for $N=10^6$ at least), the different algorithms do make a
difference to the power.  The best fit to the finite $N$ mean
field value is that using a walk of fixed length, fixed numbers of
edges and vertices added each time and a new walk started only
with every new vertex added ($v=1$) which is essentially the
original Saram\"aki-Kaski algorithm, as figure
\ref{tr3e1000000s1k2v1gamma} shows.  This has a power which is
always below the large $N$ prediction of 3 but it is close to the
mean field solution.

\begin{figure}[!htb]
\begin{center}
 \scalebox{0.8}{\includegraphics{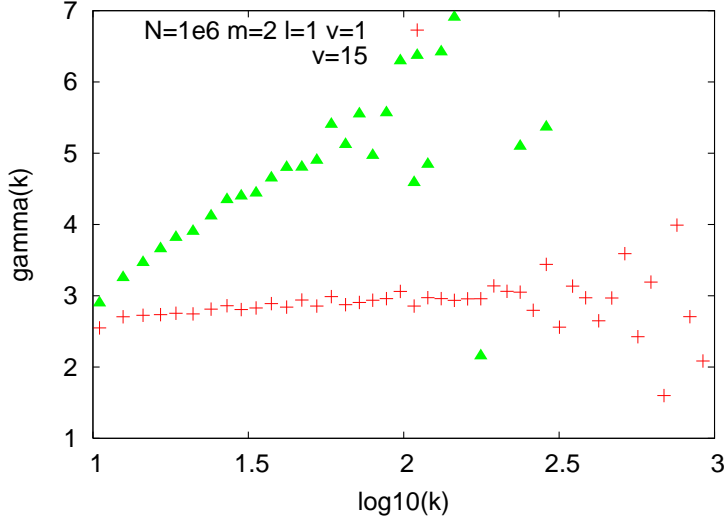}}
 \end{center}
\caption{Variation of the effective power $\gamma(k)$ for
different variants of the random walk algorithm but for walks of
average length of one step. All with $N=10^6$, $\epsilon=1$, $m=2$
and $l=1$. The $v=1$ case has a fixed length walk and is close to
the large $N$ value of $\gamma=3$.  The Markov process walk though
is expected to be similar to a mixed random/preferential
attachment algorithm with $1/2 = p_l \approx p_v$ so we expect
$\gamma=5$ in the large $N$ limit.  Indeed the $v=15$ example is
tending towards this value and is certainly has much higher
power.}
 \label{tr3e1000000s1k2v135715gamma}
\end{figure}

As was noted earlier, when a Markov process is used to choose
walks of random length (option C) this simulates a mixed
preferential attachment and random attachment algorithm. For such
cases with an average walk of length $l=1$ half the edges are
connected to a random vertex so we would expect a power of five.
Interestingly this is never quite reached so a network of a
million vertices is still not large enough though the data are
clearly tending towards this expected value, and it is certainly
bigger than the $\gamma=3$ power found when a fixed walk is used.
Figure \ref{tr3e1000000s1k2v135715gamma} shows this.

\begin{figure}[!htb]
\begin{center}
 \scalebox{0.6}{\includegraphics{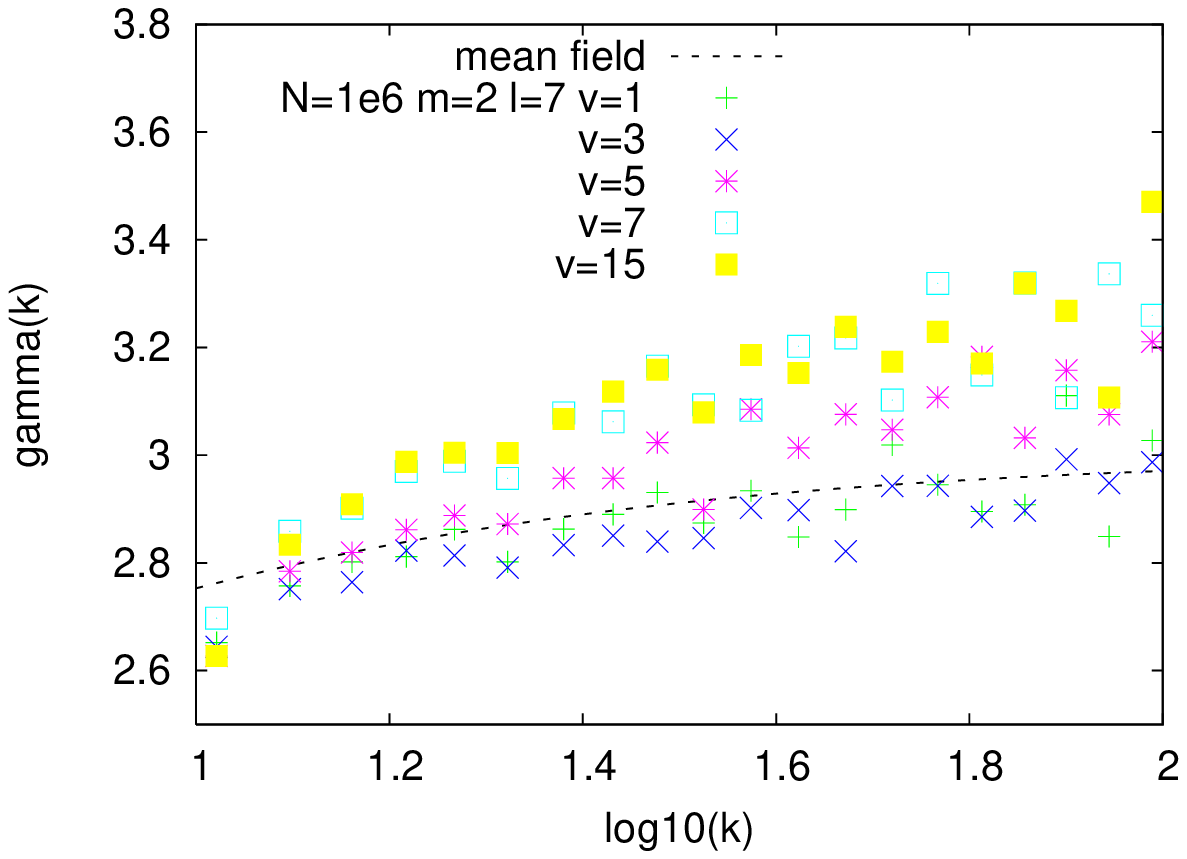}}
 \scalebox{0.6}{\includegraphics{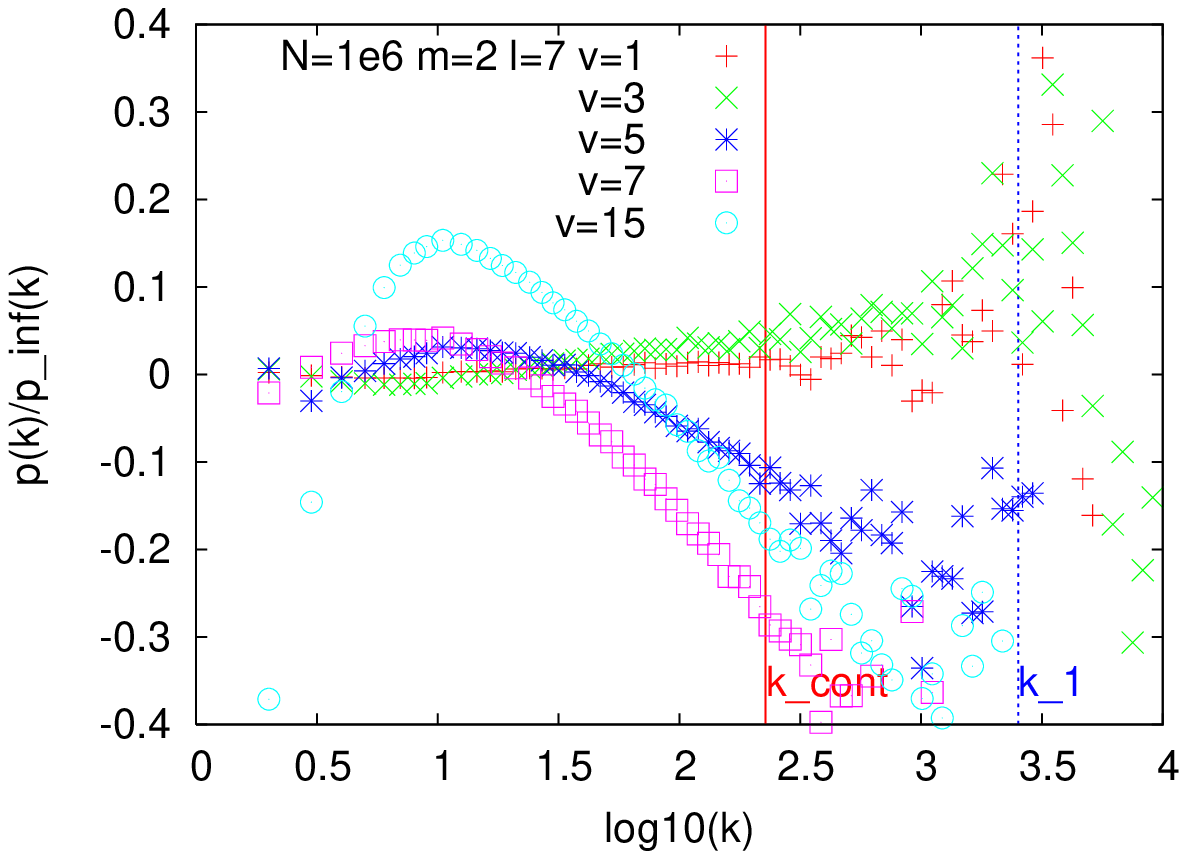}}
 \end{center}
\caption{Variation of the power law behaviour for long walks with
different variants of the random walk algorithm.  All with
$N=10^6$, $\epsilon=1$, $m=2$ and $l=7$. On the left its the
effective power with the straight line for the corresponding
numerical mean field solution. On the right the deviation from the
large $N$ mean field solution.}
 \label{gammam2l7v135715}
\end{figure}

On the other hand, other variations of the walk algorithm, even
for long walks, $l=7$, while equally well approximated by power
laws, have powers which can be consistently ten or twenty percent
higher than the finite $N$ mean-field solution as figure
\ref{gammam2l7v135715} shows.
This effect mitigates the finite $N$ reduction in the effective
power as compared to the large $N$ mean field prediction (here
3.0).  It is clear from this that while changes in the random walk
algorithm and parameters do not alter the shape of the
distribution from one that is roughly approximated by a power-law,
it does produce differences in the measured powers.

\begin{figure}[!htb] 
\begin{center}
 \scalebox{0.8}{\includegraphics{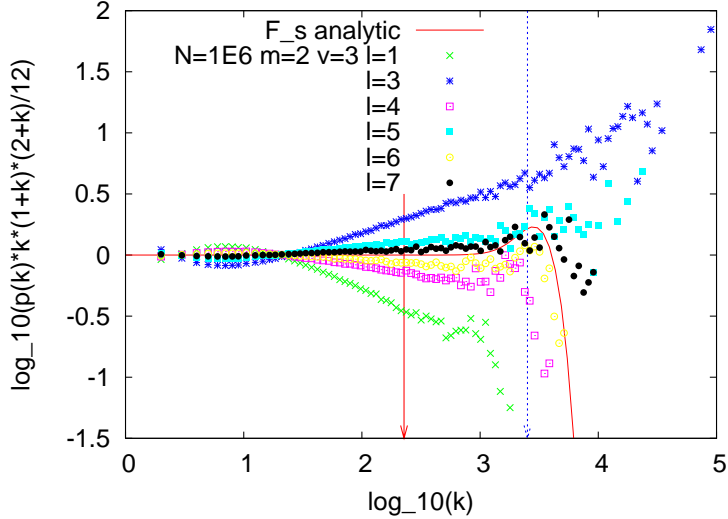}}
 \end{center}
\caption{Data is for random walk algorithms starting a new walk
from a new random vertex for every edge added, making a fixed
length walk ($v=3$), creating graphs of average degree 4 ($m=2$)
and $N=10^6$ vertices.  The length of the walk is varied from
$l=1$ to $l=7$.  Data is the average of 100 runs.  Note that again
there is clear evidence of good power law behaviour even for the
short walks.  However there is significant deviation from the form
of the mean field solution for short walks, which decreases for
longer walks.  Also note evidence of some finite size features
similar to $F_s$ for large degrees $k \sim 1000$.  The mean field
solution for the equivalent graph is the continuous line in the
centre. The mean field calculated values for $k_\mathrm{cont}$
(left) and $k_1$ (right) are indicated by the vertical lines.}
 \label{ftr3scaling}
\end{figure}

As noted the large $N$ corrections occur at high degrees $k \sim
k_1$ where the data is poor anyway, for all practical purposes we
may as well compare against the long time mean field solution
$p_\infty(k)$ of \tref{ddmfinfsol}. This is done in figure
\ref{gammam2l7v135715} for varying $\mathtt{v}$ and in figure
\ref{ftr3scaling} for varying $l$.
Again the evidence for power law behaviour is clear from even the
shortest walks, but only the longer ones come close to the exact
mean field form expected for graphs of this type.  Walks which
contain some zero length walks ($v=7$ and $v=15$) show larger
deviation reflecting the way they mimic mixed preferential and
random attachment.

Overall we see that the appearance of a long tail and scale-free
behaviour is a robust result of all non-trivial walk algorithms.
This is presumably because the relevant scale is a correlation
distance for the degree of vertices $\xi$ steps apart, and it
appears that $\xi \lesssim 1$.\tnote{Can we measure such a
correlation function and length?} However the power of the
distribution is varies considerably and is sensitive to the
details of the algorithm.

\subsection{Global length scales}

\begin{figure}[htb]
\begin{center}
 \scalebox{0.8}{\includegraphics{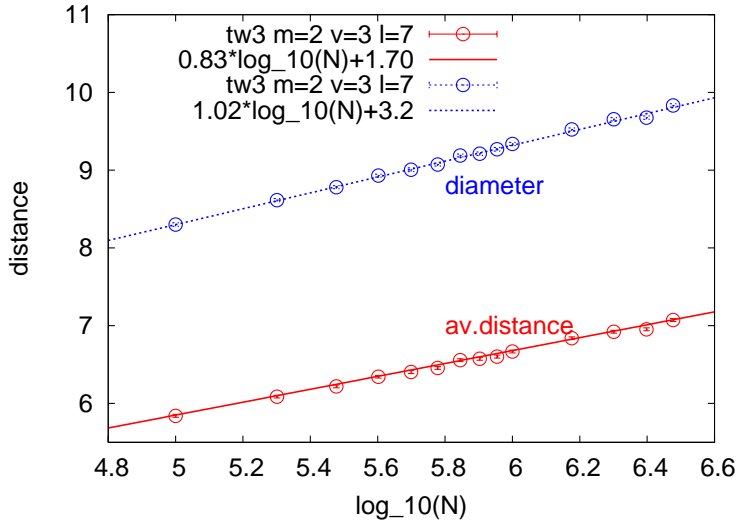}}
 \end{center}
\caption{Average shortest distances and diameters for different
total numbers of vertices $N$, with the average degree held fixed
($m=2$).  The error bars on data points are drawn but are
comparable with the size of the symbol. The data are for 100 runs
a new random walk starting for every edge added (two per new
vertex) and of fixed length $l=7$ ($v=3$).  The straight lines are
a best fit to the data.}
 \label{fdistvaryNb}
\end{figure}

The diameter and average shortest path length were not studied in
\cite{SK04}.  We note that in our random walk algorithm they show
the expected behaviour of scaling as $\ln(N)$ as figure
\ref{fdistvaryNb} shows.\tnote{Comes from the version 3 of
\texttt{timwalk} programme.} The average shortest distances
between points and the diameters (a lower bound at least) are
shown for different total numbers of vertices $N$, with the
average degree held fixed ($m=2$) and a walk length of seven
($l=7$) for an exemplary algorithm. Both clearly scale with
$\ln(N)$.
Other variations of the walk algorithm show similar behaviour
though the diameters and shortest distance measures do depend on
the particular random walk algorithm used.

The next figure \ref{fdistvarys}\tnote{From version 3 of
\texttt{timwalk}.} shows how average shortest distances and the
diameters are vary for different fixed numbers of vertices
$N=10^6$, fixed average degree $K=4$ but varying length for the
random walk. Just as in the case of the clustering coefficient
\cite{SK04} there is an interesting pattern for odd and even walk
lengths when the walks are of fixed length (here the \texttt{v=3}
runs). This is an artifact of the discrete nature of the algorithm
because there is a good chance on short walks that one returns to
the original vertex when the length of the walk is even. It is not
seen in the smoother algorithm of the \texttt{v=15} runs where the
number of edges added and the number of steps taken is varied but
the averages are kept the same. As the walk lengthens we are
tending to a fixed value suggesting that the simplest algorithms
generate some correlations for short walks.
\begin{figure}[htb]
\begin{center}
 \scalebox{0.8}{\includegraphics{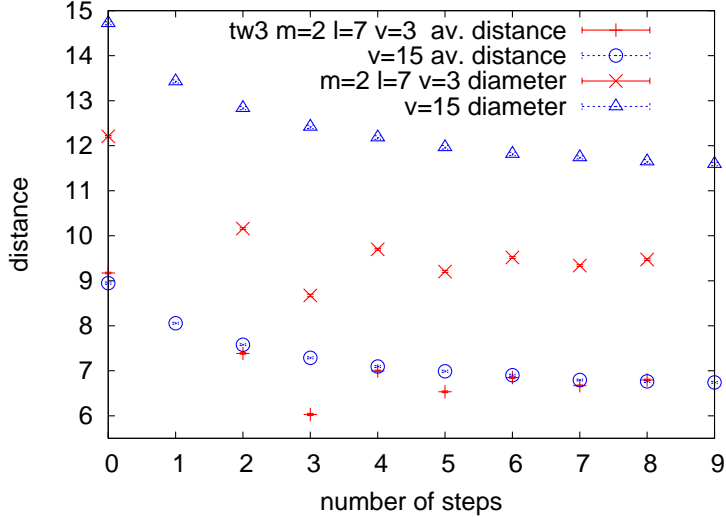}}
 \end{center}
\caption{Average shortest distances and diameters for varying
lengths of random walk, fixed vertex and edge numbers ($N=10^6$,
$\epsilon=1$, $m=2$) with walks starting from a random vertex. The
data shown are for two types of algorithm. Crosses are for fixed
walk length starting a new walk for every edge ($v=3$). The
circles and triangles have a variable number of edges added per
vertex and a new walk of variable length is used for every new
edge but averages are kept as before ($v=15$). Note the dependence
on the odd/even nature of the $v=3$ case and the clear trend as
the walk length gets longer.  Error bars are shown but are smaller
than the sizes of the symbols.}
 \label{fdistvarys}
\end{figure}

\section{Weighted Graphs}\tnote{Need to unify notation: (Tim = Jari)
$m = m$, $l= l_w$, Tim uses $s$ for step length but its naturally
used for strength.  Jari uses $l_w$ for the degree walk when $w$
could be for weight yet $l_d$ is for the weighted walk when it
could stand for degree.  The hard bit is changing the diagrams.}

Many graphs are not simple graphs but their vertices and edges
often carry other information.  This is readily taken into account
by considering the edges to be \emph{weighted}
\cite{YookWeightedSF,BBV04b,BBV,NewmanWeighted2004,Onnela2004}, so
that every edge is characterised by its weight $w$. Then, a
natural generalisation of vertex degree is the vertex
\emph{strength} $s$ \cite{BBV04b}, defined as the sum of weights
of edges connected to the vertex. The weights provide an
additional degree of freedom, and their dynamics can be coupled to
network evolution. Recently, BBV (Barrat \emph{et al.}) \cite{BBV}
proposed an algorithm where networks are grown based on a
strength-driven preferential attachment rule. In the BBV model,
new nodes joining the network are connected to vertices chosen
with a probability proportional to their strength with links
initially having unity weight. Then, an amount of $\delta^\ast$ of
extra weight is divided among the old edges of each parent
vertex in proportion to their weights: $w_{ij} \rightarrow w_{ij}+\delta^\ast w_{ij}/s_i$.
This leads to asymptotic power-law distributions of both
the vertex degrees and the vertex strengths, with an exponent
$\gamma=\left(4\delta^\ast+3\right)/\left(2\delta^\ast+1\right)$,
i.e. the power law gets broader with increasing $\delta^\ast$.
Also the distribution of weights follows an asymptotic power law,
$P(w)\sim w^{-\alpha}$, where $\alpha=2+1/\delta^\ast$.

In the following, we will show that the walk algorithm can readily
be generalised to the weighted case, providing a natural model for
evolving weighted networks. We will focus just on the weight
aspect of the problem and work in this section with a basic random
walk algorithm, so that we always use walks of fixed lengths and
at every time step add one vertex ($\epsilon=1$) and add a fixed
number of edges $m$, each attached at one end to the new vertex.

The algorithm we use is as follows. The network dynamics is
divided into two aspects: i) network growth and ii) modification
of the existing weights, which both take place successively during
each time step $t$.  Both cases are based on random walks, where
we modify the random walking rule so that the next step in the
walk is always chosen so that the probability of following a link
is directly proportional to its weight, i.e. if the walker is
located at vertex $v_i$, it next moves to vertex $v_j$ with the
probability $w_{ij}/\sum_k w_{ik}$, where the sum is over all
neighbours of $v_i$.

With the exception of the above modification, the network growth phase proceeds
as detailed earlier, so that the $m$ vertices are chosen using random walks of
length $l$. If we assume that there is no correlation between the
strength of neighbouring vertices, this reduces to the simple case
of
\beq
 \Pi = s/S(t),
  \label{Pisdef}
\eeq
that is, we will have pure preferential attachment in terms of
strength rather than degree. When the parent vertices have
been selected, an initial weight of $w_0$ is assigned to the new edges.
Then, we modify the existing weights by performing a second type of walk
so that
\begin{enumerate}
\item To start the random walk we choose a vertex $v_j$ in the existing graph,
$G(t)$, choosing at random from a uniform distribution.
\item Now make one step in a random walk on the graph by choosing
one of the neighbours of $v_j$ at random using the above biasing rule. The edge we follow has
its weight increased by $\delta$.
\item Repeat the previous step $l_d$ times.
\end{enumerate}

The strength distribution in the mean field approximation follows a
similar equation as for the degree, namely
\begin{eqnarray}
n(s,t+1) - n(s,t)
 &=& r_s
   [-n(s,t)\Pi(s,t) + n(s-\delta,t) \Pi(s-\delta,t) ]
   \nnel
   && + \epsilon \delta_{s,w_0} .
\label{seqn}
 \\
 r_s &:=& [ 2 l_d  + (w_0 / \delta) ], \label{rsdef}
\end{eqnarray}
The total strength $S(t)$ is given by
\beq
 S(t) = \sum s n(s,t) = S(0) + 2(l_d \delta + w_0)
\eeq
while now $N(t) = N(0) + t$. The analysis of the strength distribution is then exactly as
before, and for large graphs
we find that the asymptotic form for the distribution is a power law
\bea
 \lim_{s \ra \infty } \lim_{t \ra \infty}n(s,t) &=& s^{-\gamma_s},
 \\
\gamma_s &=& \frac{3m+4l_d\delta}{m+2l_d\delta}.
 \label{gsdef}
\eea

Note the relation to the BBV model's exponent for the strength distribution \cite{BBV},
$\gamma_{BBV}=(4\delta^\ast+3)/(2\delta^\ast+1)$.
The total increase of weight in the modification phase equals
$\Delta=m\delta^\ast$ in the BBV model, and $\Delta=l_d\delta$ in our weighted walker model.
Both exponents can be rewritten using this quantity as $\gamma=(3m+4\Delta)/(m+2\Delta)$.

Now, we may expect that for individual vertices $k \propto s$,
because in the network growth phase the probability that a random walk arrives at a given vertex is
proportional to its strength. Substituting this as an ansatz we
find that the degree distribution also follows a power law with
$\gamma_k=\gamma_s$.
Note that the same the exponents also emerge from analysis based on continuum
mean-field rate equations in the same manner as done in Ref.~\cite{BBV}.

It is also possible to apply the mean field approach to the weights on
each edge.  In the limit of $N \rightarrow \infty$, $t \rightarrow \infty$
we
again find a power law for the distribution of weights of
\begin{equation}
p(w) \propto w^{-\alpha}, \label{alphadef}
\end{equation}
with the exponent $\alpha = 2+m/(l_d\delta)$. This also reproduces the form found in \cite{BBV}.

We can conclude that the main characteristic distributions of
networks grown with the weighted walker model are equivalent to
the ones of the BBV model. However, the models are not identical.
We have deliberately chosen to start the weight modification walks
from randomly selected vertices, instead of ones connected to
newly joined vertices. This illustrates that the distributions are
of a general nature and a result of strength-driven attachment in
combination with preferential increase of weights -- strong
weights get stronger, a feature that is implicitly present in the
BBV model in the form of dividing the weight increase
proportionally among edges. Furthermore, as shown for unweighted
networks elsewhere in this paper and in Ref.~\cite{SK04}, we
expect other characteristics such as the degree of clustering and
the network diameter to depend on the random walk lengths.
Especially, with short growth-phase random walk lengths $l$, the
networks are expected to show high degrees of clustering, a
feature found in several real-world networks. We choose to leave
further investigations of these issues for future work.

\subsection{Numerical Results}

\begin{figure}[t]
\begin{center}
\includegraphics[width=0.6\textwidth]{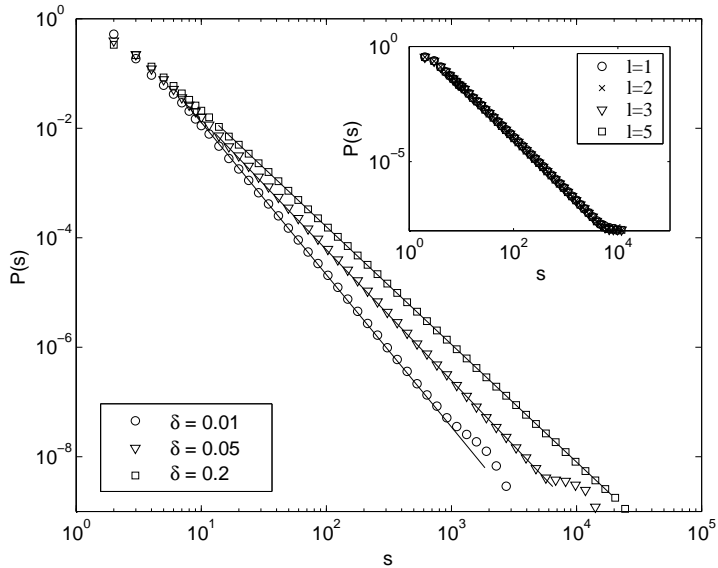}
\end{center}
\caption{Distribution of vertex strength $p(s)$, averaged over
$1,000$ realisations of $N=2\times 10^5$, $m=2$ networks grown
using the weighted walk algorithm with $l=15$, $l_d=30$, and
$\delta=0.01$ ($\circ$), $\delta=0.05$ ($\triangledown$) and
$\delta=0.2$ ($\square$). The solid lines indicate slopes for
respective asymptotic
 power laws calculated using \tref{seqn}. Inset: $p(s)$
averaged over $2,500$ realisations of $N=5\times10^4$ networks,
with $\delta=0.1$, for various walk lengths $l=1,2,3,5$. The
power-law behaviour is visible even for the shortest walks.}
\label{dist1}
\end{figure}

Figure \ref{dist1} illustrates the probability distribution for
strength $p(s)$ calculated from simulating the random walker
network growth process, together with the mean-field prediction of
\tref{seqn}. The networks were grown to size $N=2\times 10^5$,
with $l=15$, $l_d=30$, $m=4$ and $\delta$ as illustrated. The
results are averages over $1,000$ realisations.  They fit the mean
field power laws of the form \tref{gsdef} as figure \ref{dist1}
shows.

\begin{figure}
\begin{center}
\includegraphics[width=0.6\textwidth]{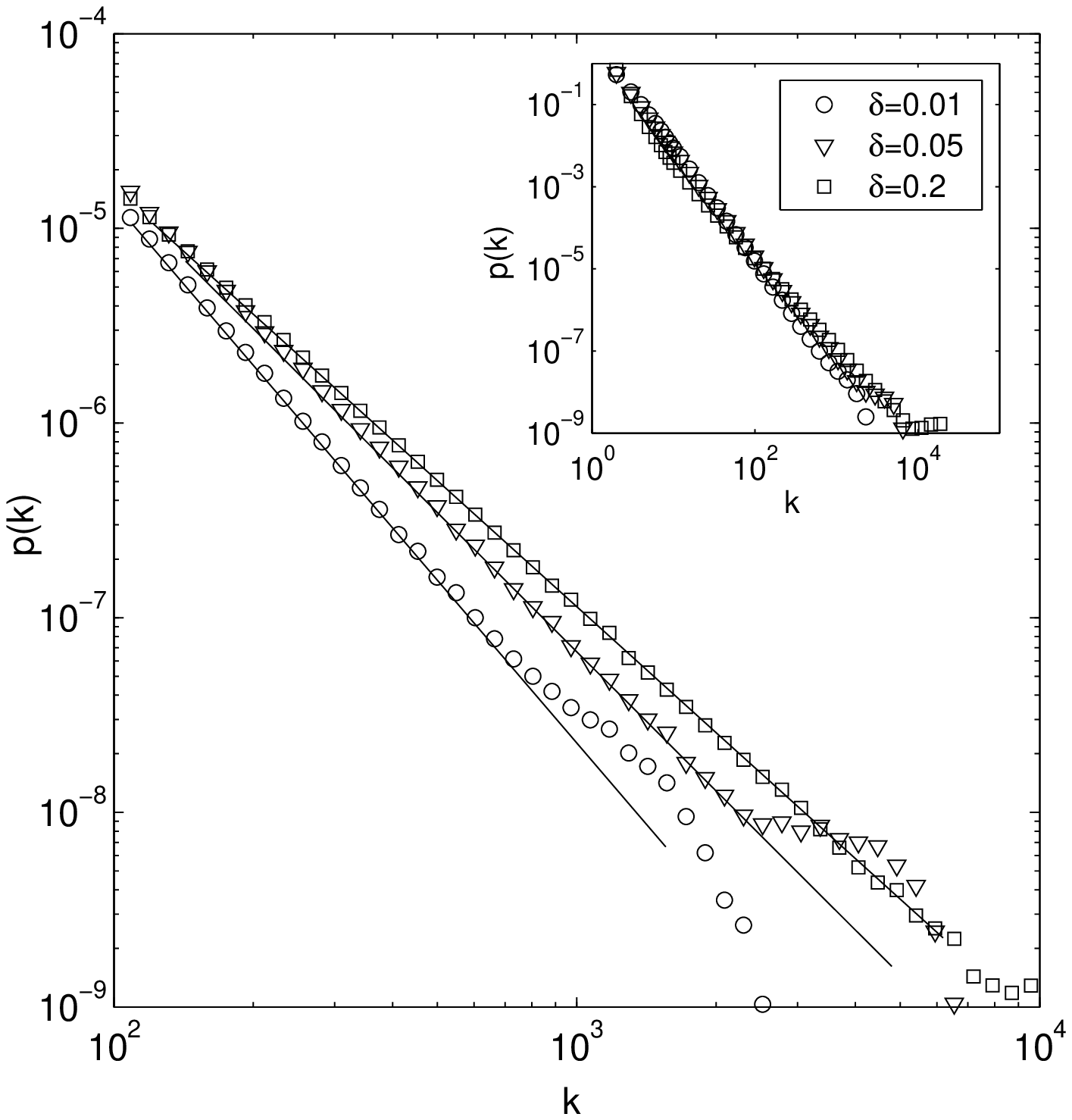}
\end{center} \caption{Degree distribution $p(k)$ for the same
networks as in figure \ref{dist1}. The solid lines indicate slopes
for mean-field power laws. The inset shows the distribution over
the whole $k$ range.} \label{dist2}
\end{figure}

\begin{figure}
\begin{center}
\includegraphics[width=0.6\textwidth]{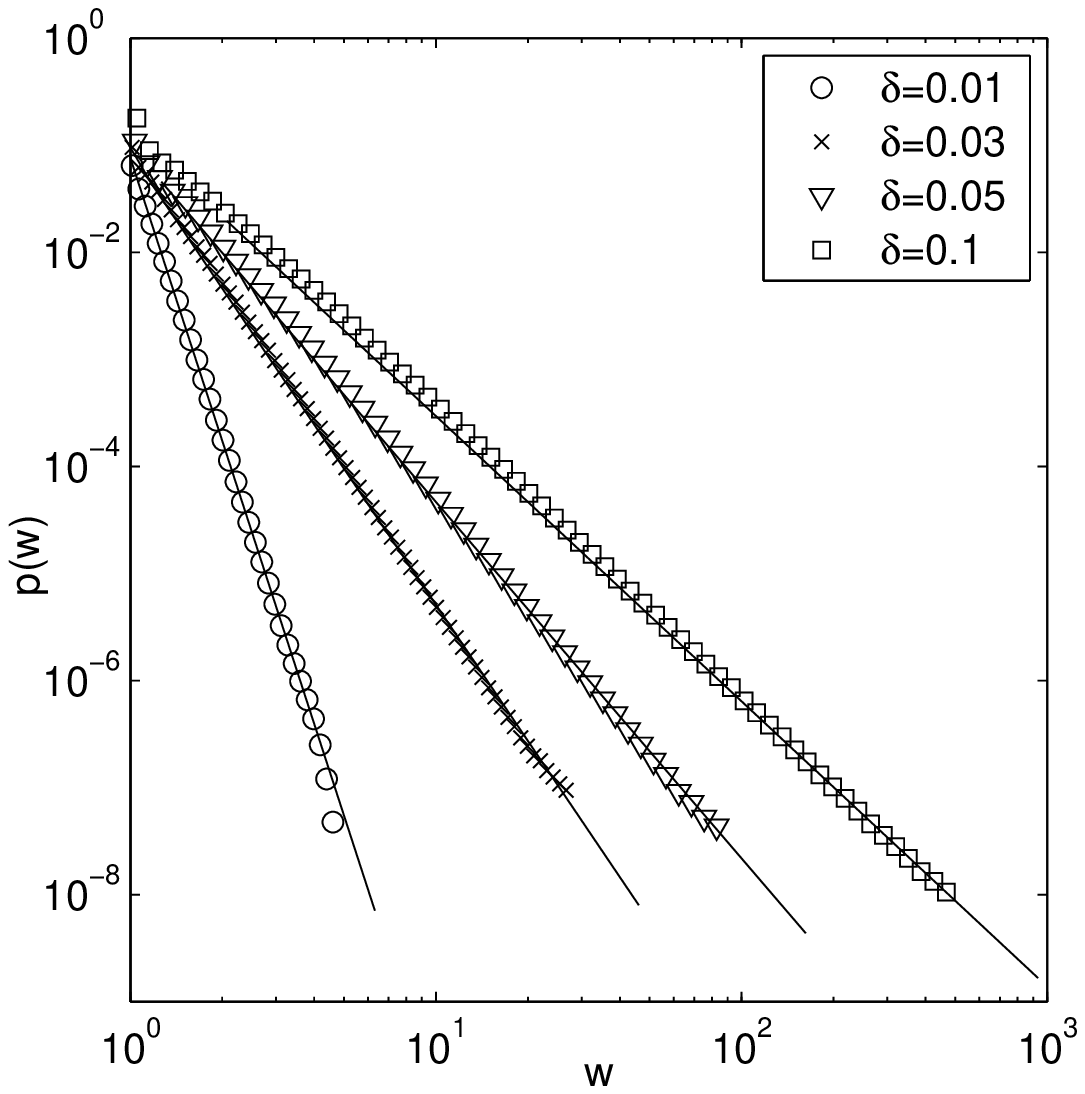}
\end{center} \caption{Weight distribution $p(w)$ for $N=2\times
10^5$ networks, averaged over $10^3$ realisations, with $m=2$,
$l=15$, $l_d=30$ and $\delta$ as shown in the legend. The solid
lines indicate slopes for mean-field power laws of
\ref{alphadef}.}\label{weights}
\end{figure}

As noted, we expect in this algorithm that the degree distribution
in this weighted random walk algorithm to show the same form as
the strengths and this is seen in figure \ref{dist2}. Finally,
figure \ref{weights} illustrates the power law distribution
of weights. Also in this case the slopes match the mean-field
approximation of \ref{alphadef}.

\section{Conclusions}

Random walks on graphs provide a variety of different types of
network, as seen in the variations in distance
scales\footnote{Also for clustering coefficients as seen in
\cite{SK04}.}. However, apart from some special cases in the limit
of zero length (no) walks, they are invariably characterised by
having a degree distribution with a very long tail, and a power
law will often be a sufficiently good description of this tail.

We have stressed that most networks in numerical studies or in
studies of real systems are mesoscopic systems. That is even for
systems of the order of a million vertices, finite size effects
are noticeable. For instance a simple power law fit to data from a
theoretical model should give a power that is anywhere from 0.1\%
to 10\% below that expected for the infinite sized graph due to
the effects of small degree deviations from simple power laws.
Further our numerical studies are idealised with 100 or 1000
examples used so we expect real noisy single data sets will be
harder to interpret. Note also that such differences from an exact
power law are hard to detect by eye on log-log plots of
distributions, even in our idealised situations. Thus while
power-laws reported in the literature may be an `acceptable'
description of a data set in many circumstances, it may be
difficult to distinguish between different underlying processes or
even between different types of degree distribution \cite{Mit04}.

However, given that proviso, we believe that the a random walk
algorithm does provide one of the few realistic explanations why so many
different systems have degree distributions which are consistent
with power-laws.  Further we suggest that many of these real world
networks are in fact genuine scale-free networks and would have
pure power laws in the infinite time, infinite graph limit. We
have studied a wide set of variations on the basic random walk
algorithm of Saram\"aki and Kaski \cite{SK04}, including an
extension to more realistic weighted graphs. In almost all cases
we have found power laws emerge naturally. Various powers for the
power law are possible depending on the algorithm and on its
parameters but a power-law like distribution is an extremely
robust result of the generic random walk algorithm. The random
walk algorithm exploits the structure of the
graph\footnote{Indeed, as far as the degree distribution goes, a
simple preferential attachment model need have no graph present at
all. For instance Simon \cite{Simon55} makes no reference to a
graph though one can invent one if one wishes for his examples.
Conversely, while the web provides a natural example of a network,
one can easily count the links on a web page and hence obtain the
outgoing degree distribution without any reference to the
underlying graph. The preferential attachment model of Barab\'asi
and Albert \cite{BA}, who refer to the graph, is no different to
that of Simon \cite{Simon55}, who has no graphs, as far as the
degree distribution goes.} yet it requires no global information
to operate. This in sharp contrast with most numerical and
algebraic analyses, for example \cite{Simon55,BA,BBV}, where
preferential attachment is assumed and implicit global information
is used in the normalisation. Thus in this sense we see the random
walk algorithm as a process of self-organisation, the very
structure of the graph inevitably leads microscopic local
processes to a scale-free form.

While this may be a useful way to understand why so many
scale-free networks are seen in the real world, the walk algorithm
could be a useful in practical problems.  Due to its robustness
and purely local nature, the random walk algorithm could  be used
to engineer new networks which self-organise to a scale-free form.
For instance this might be of use for  distributed computing and
peer-to-peer network problems.

\section*{Acknowledgements}

We would like to thank the organisers of CNET 2004 for providing
the opportunity to work on this problem and S.Redner for useful
comments on an earlier version. TSE would like to thank S.Klauke
for early discussions about the random walk algorithm \cite{Kla},
D.Hook and R.J.Rivers for continuing discussions, and the ISCOM
network for providing a productive and broad forum. JPS is
supported by the Academy of Finland, Project No.~1169043 (Finnish
Centre of Excellence programme 2000-2005).


\newpage

\appendix

\section{Mean Field Finite Size Calculations}

The mean-field equation for the degree distribution for a network
grown with mixed random and preferential attachment was given in
\tref{neqn}. In the long time, large graph limit for pure
preferential attachment (corresponding to our parameters
$p_v=0,\epsilon=1$) the solution is $p(k,t) = p_\infty(k)F_s(k,t)$
\tref{ddmfsol} where the finite size corrections to the infinite
time distribution $p_\infty$ are contained in the function $F_s$.
A solution for the case where the average degree of the network
tends to two ($m=2$ here) was given by Krapivsky and Redner
\cite{KR02} (see also \cite{DMS01,KK00,ZM00}). We have followed
the approach of \cite{KR02} and generalised this to arbitrary $m$.
We define a generating functional
\beq
 F =
 \sum_{t=1}^{\infty}\sum_{k=m}^{\infty}
 w^{t-1} z^k n(k,t)
 \label{Ftildef}
 \eeq
 Switching to variables $x$ and $y$ where
 \beq
 x = - \frac{1}{4} \ln(1-w) + \half \ln\left(\frac{z}{1-z}\right)
, \qquad
 y = - \frac{1}{4} \ln(1-w) - \half \ln\left(\frac{z}{1-z}\right)
\eeq
the mean field equation \tref{neqn} becomes
\beq
 \half \frac{\partial F}{\partial x} -F = \frac{z^m}{(1-w)^2}
\eeq
which has the solution
\beq
 F = 2 e^{2x} e^{4y} \int^x dx^\prime
 \frac{e^{(2+m)x^\prime}}{(e^{x^\prime}+e^y)^m}
\eeq
Now one must impose some initial conditions to provide the
boundary conditions needed to find the explicit solution.  The
first vertices tend to be the largest degree vertices in the long
run and so the shape of scaling function $F_s$ is sensitive to
this choice.  We choose
\beq
 n(k=m,t=1)=2, \qquad n(k\neq m,t=1)=0
\eeq
which gives
\bea
 F(x,y) &=& 2 e^{2x} e^{4y} \int_{-y}^x dx^\prime
 \frac{e^{(2+m)x^\prime}}{(e^{x^\prime}+e^y)^m} + F_b(y)
 \\
 F_b(x,y) &=& \frac{2 e^{2x} e^{2y}}{ ( 1+e^{2y} )^m }
\eea
The integral can be performed in terms of a variable
$q=e^{x^\prime}+e^y$.

Now starting from \tref{Ftildef} we see that by substituting in
the form \tref{ddmfsol} we can show that
\bea
 \frac{\partial^3}{\partial z^3} (z^{2} F)
  &=&
 \sum_{t=1}^{\infty}\sum_{k=m}^{\infty}
 w^{t-1} z^{k-m-1} (N_0+t) 2m(m+1) F_s(k,t)
  \label{Fdefder}
\eea
where $N_0=1$ is the number of vertices at $t=0$.  Working in
terms of variables $\epsilon = e^{-2x} e^{-2y} = (1-w)$ and $\eta
= e^{y}e^{-x}= (1-z)/z$ we are interested in the limit where $w,z
\rightarrow 1$ or equivalently $\epsilon, \eta \rightarrow 0$ such
that $\eta/\epsilon^{1/2} = s$ is constant.  In  this limit we
find that the left-hand side of \tref{Fdefder} can be written as
\bea
 \frac{\partial^3}{\partial z^3} (z^{2} F)
 &=&
 \frac{1}{\epsilon^{5/2}} J_m(s),
 \\
 J_m(s) &=& 2m(m+1)
  \left[
 \sum_{n=1}^{m=2}
 \frac{1}{(1+s)^n}
 +
 \frac{m+2}{(1+s)^{m+3}}
 \right]
\eea
for $m=1,2,3,4$ and we conjecture the same for higher $m$.  The
$m=1$ value coincides with that in \cite{KR02}.

Now we look at the right-hand side of \tref{Fdefder} assume that
the scaling function is of the form $F_s=F_s(k/t^{1/2})$. We are
interested in the large degree and time effects so we can
approximate the sums by integrals from zero to infinity over the
variables $\xi = k \epsilon^{1/2}$ (for the $k$ sum) and $\tau = t
\epsilon$ (for the $t$ sum).  In the same way we can approximate
$w^t \approx e^{-\tau}$ and $z^k \approx e^{-s \xi}$ and interpret
these integrals as Laplace transforms.  In particular the
right-hand side of \tref{Fdefder} is the Laplace transform over
$\xi$ (or $k$) of a function $\Phi$ where
\beq
\Phi(\xi) = 2m(m+1) \int_0^\infty d\tau \; \tau e^{-\tau}
F_s(\xi/\tau^{1/2}) .
 \label{Phidef}
\eeq
Thus \tref{Fdefder} can now be expressed as the inverse Laplace
transform
\bea
\Phi(\xi) &=& \frac{1}{2\pi i }
  \int_{c-i \infty}^{c+i \infty}
  e^{\xi s} J_m(s)
  \\
  &=&
  2m(m+1) e^{-\xi}
  \left[
  \sum_{n=0}^{m+1} \frac{\xi^n}{n!}
  +  \frac{\xi^{m+2}}{(m+1)!}
  \right]
\eea
Comparing this with \tref{Phidef} we have
\beq
  \Phi(\xi) =
  2m(m+1) \xi^4 \int_0^\infty d\zeta \; e^{-\xi^2 \zeta}
  \left(\zeta F_s(\zeta^{-1/2}) \right)
\eeq
where $\zeta= \tau/\eta^2$.  By treating this as the Laplace
transform in $\zeta$ of a function $G(\zeta)= \zeta
F_s(\zeta^{-1/2})$ with respect to a variable $p=\xi^2$ we just
have to use inverse standard Laplace transforms to produce the
answer \tref{Fsdef}.

\section{Supplementary Material}

These are provided for information and will not be in the journal
version.

\subsection{Degree distributions for even $v$ algorithms}

\begin{figure}[!htb]
\begin{center}
 \scalebox{0.5}{\includegraphics{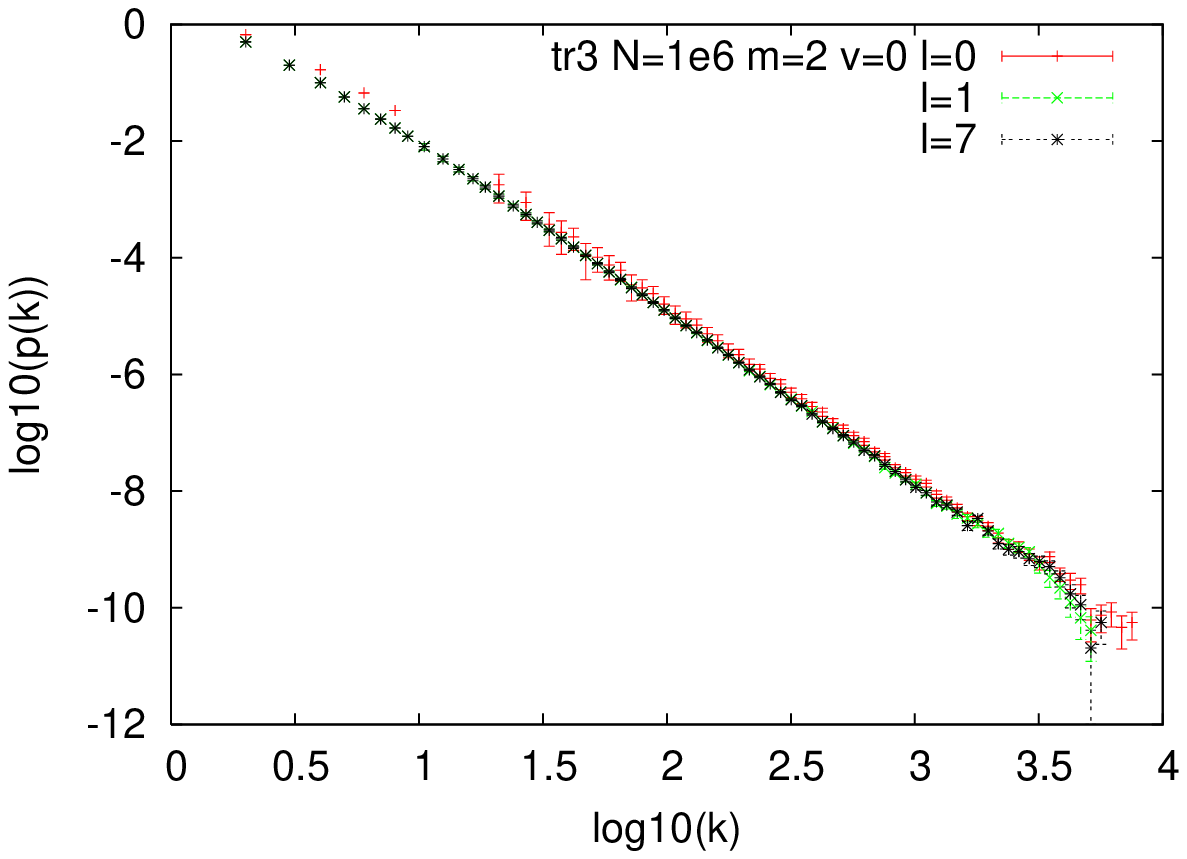}}
 \scalebox{0.5}{\includegraphics{tr3e1000000s017k2v2r_logbin_av.eps}}
\\
 \scalebox{0.5}{\includegraphics{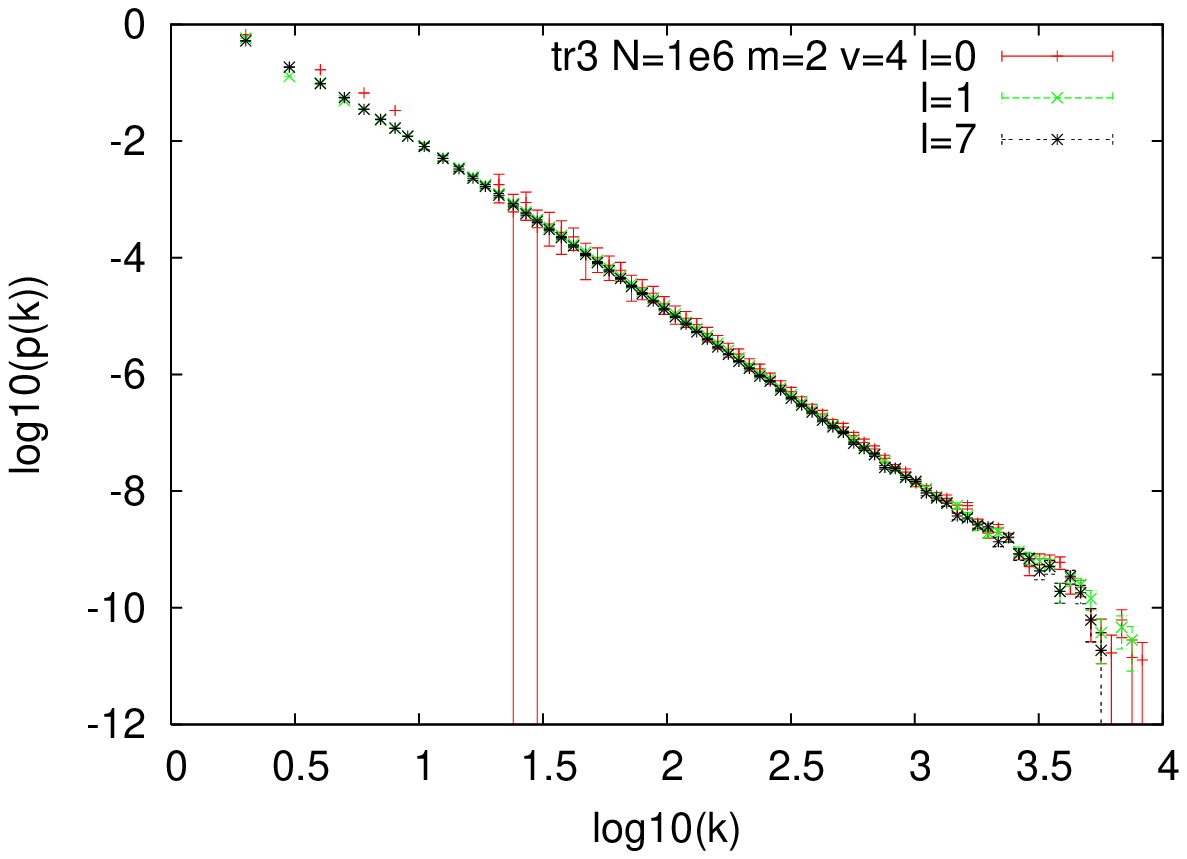}}
 \scalebox{0.5}{\includegraphics{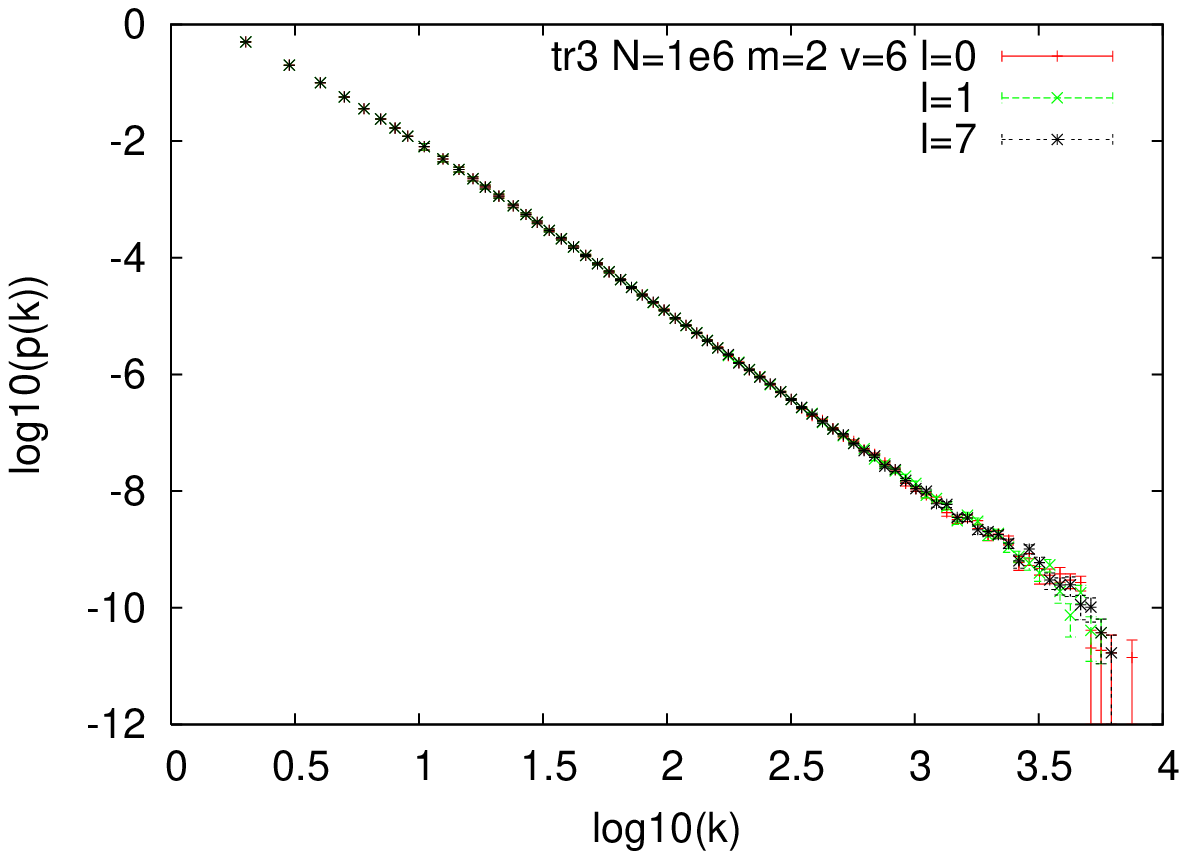}}
\end{center}
\caption{Plots of $\log_\mathrm{10}(n(k))$ vs
$\log_{\mathrm{10}}(k)$ for walks started from a random end of a
randomly chosen edge.  All with one vertex ($\epsilon=1$) and two
edges ($m=2$) added at each time step and a total of $10^6$
vertices added. In each graph the results are shown for average
walk lengths, $l=s$, of 0,1 and 7 steps with data averaged over
100 runs and the data are binned with bins chosen such that
$k_\mathrm{max}/k_\mathrm{min} \approx 1.1$. On the left runs have
fixed walk length while on the right a random length is chosen
using a Markov process.  The top row does one per at each time
step, while the bottom row starts a new one for each edge added.
Multiple edges are allowed here.}
 \label{ftr3n1e6s017v0246}
\end{figure}

Figure \ref{ftr3n1e6s017v0246} shows the algorithms which have
walks of various lengths starting from a random end of a randomly
chosen edge.

\subsection{Semi log plots}

\begin{figure}[!htb]
\begin{center}
 \scalebox{0.5}{\includegraphics{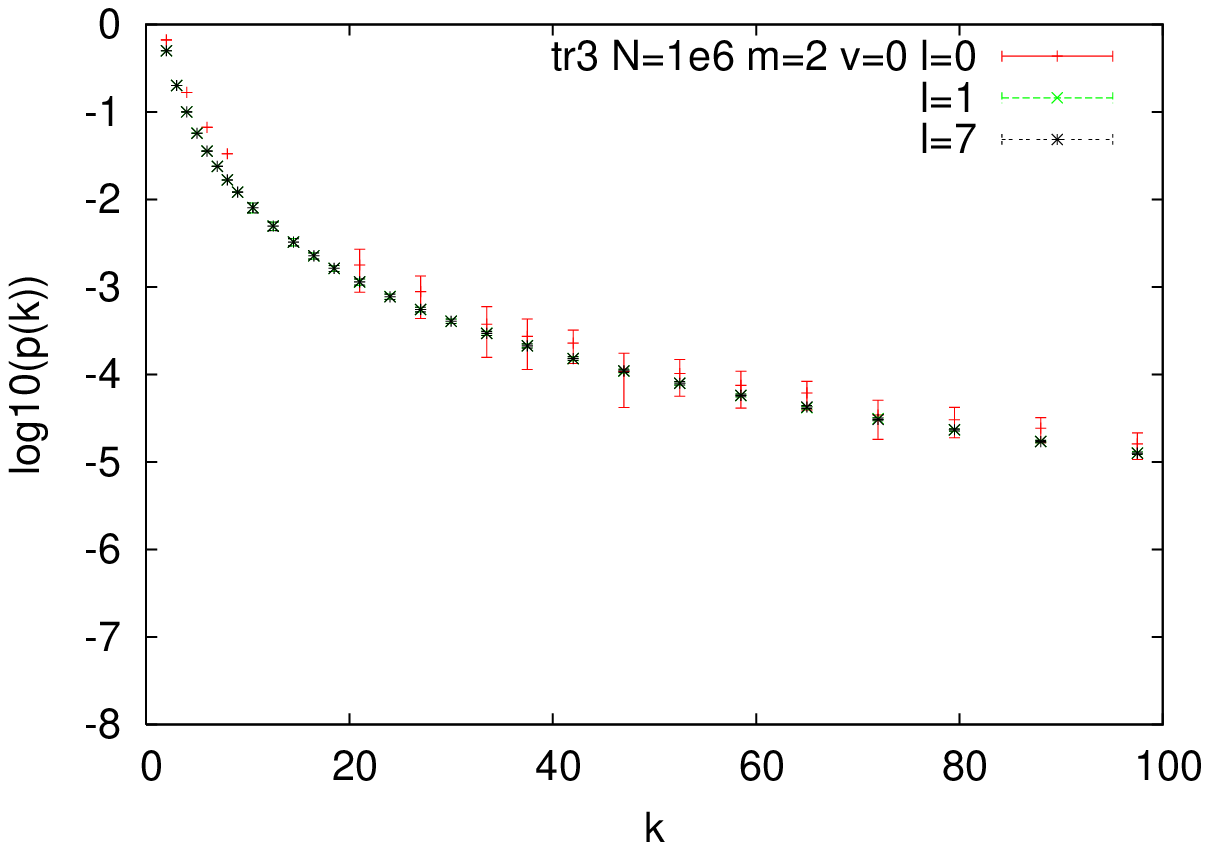}}
 \scalebox{0.5}{\includegraphics{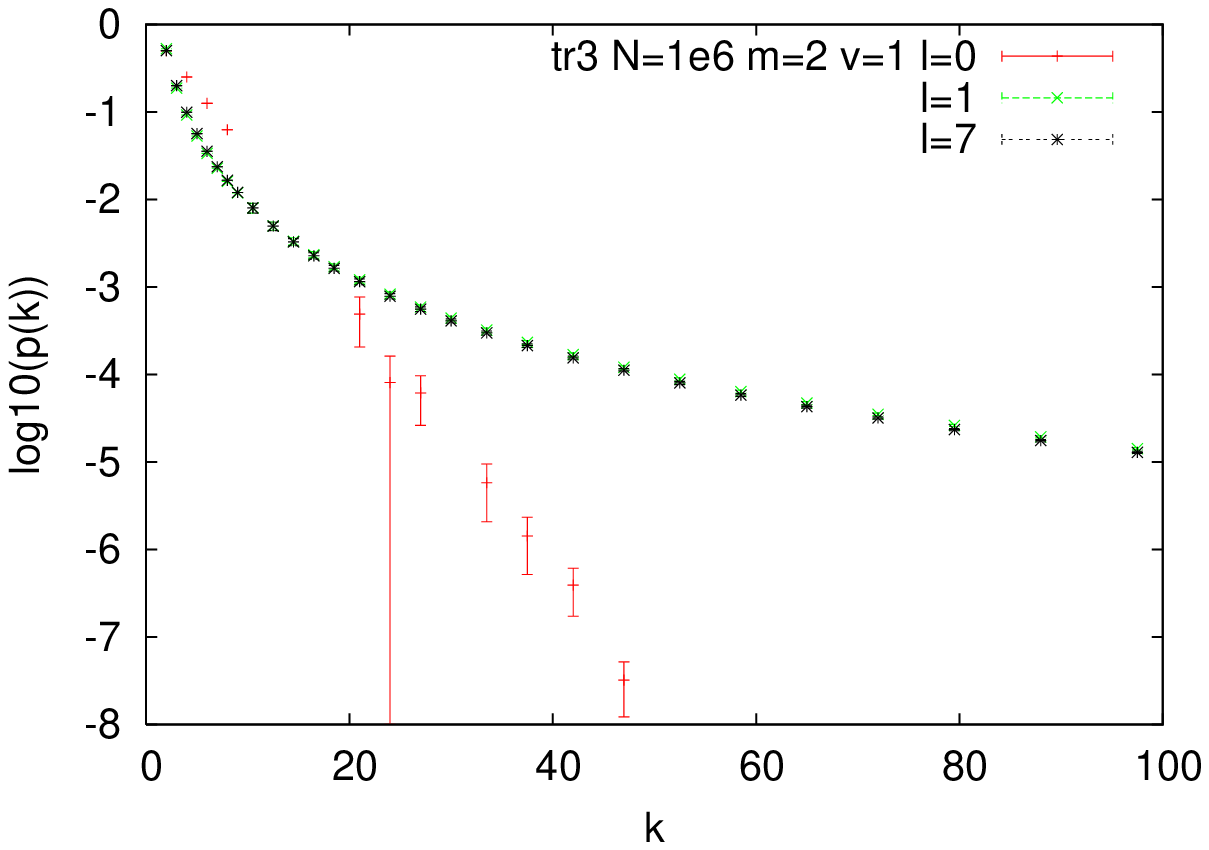}}
\\
 \scalebox{0.5}{\includegraphics{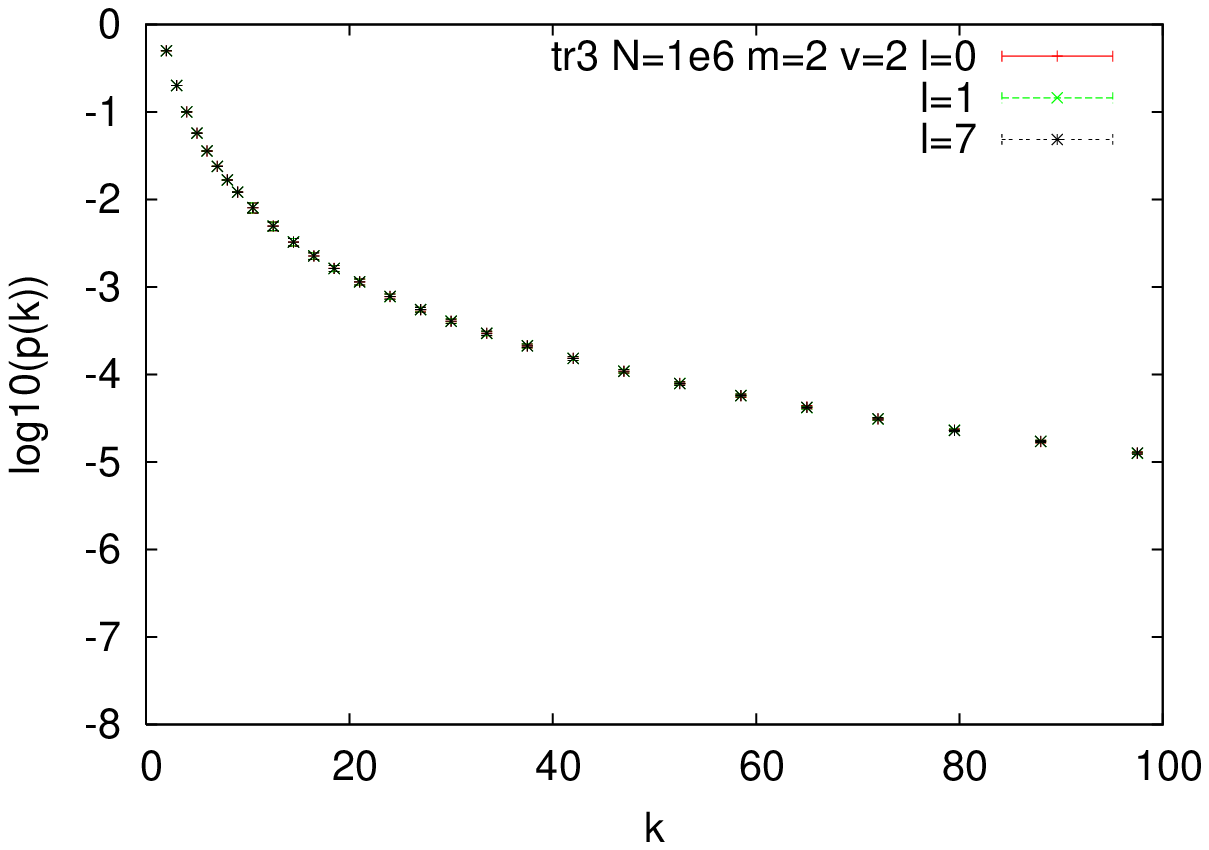}}
 \scalebox{0.5}{\includegraphics{tr3e1000000s017k2v3r_logbin_av_sl.eps}}
 \\
 \scalebox{0.5}{\includegraphics{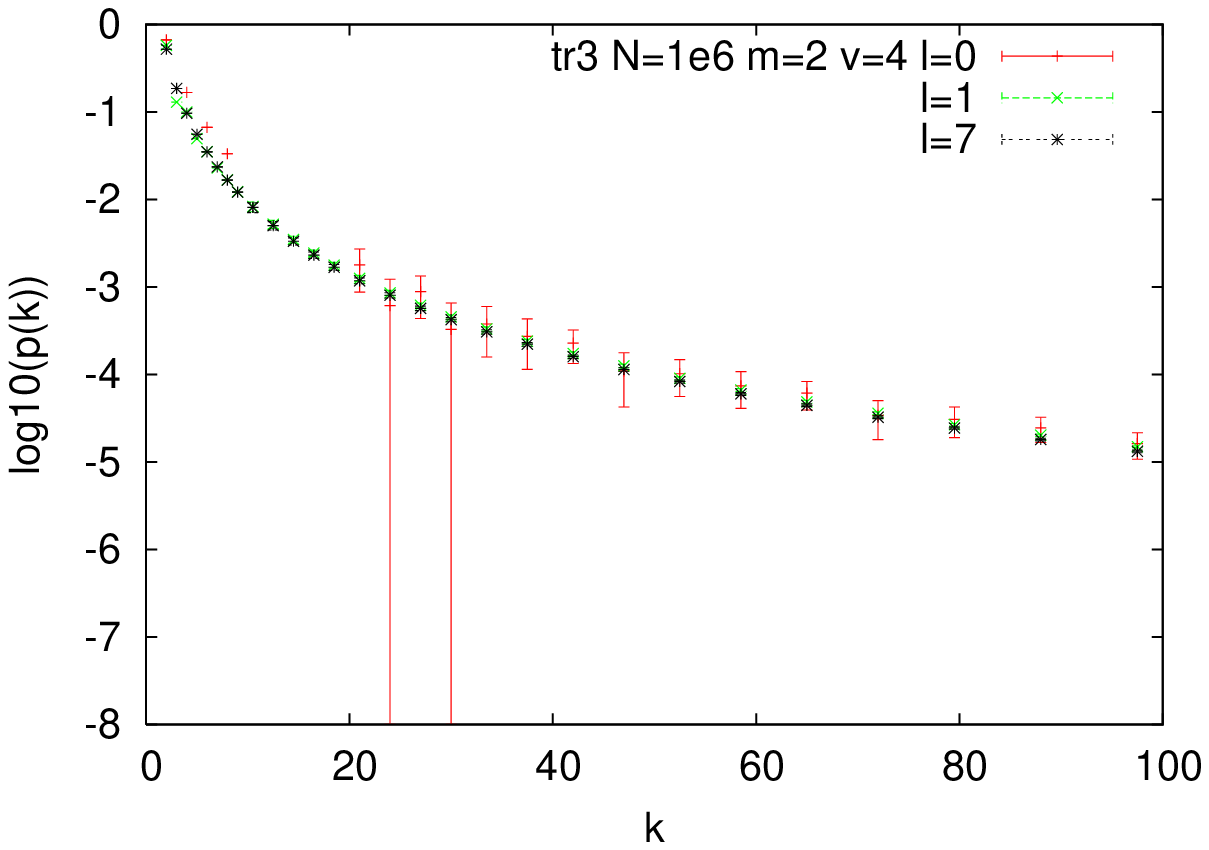}}
 \scalebox{0.5}{\includegraphics{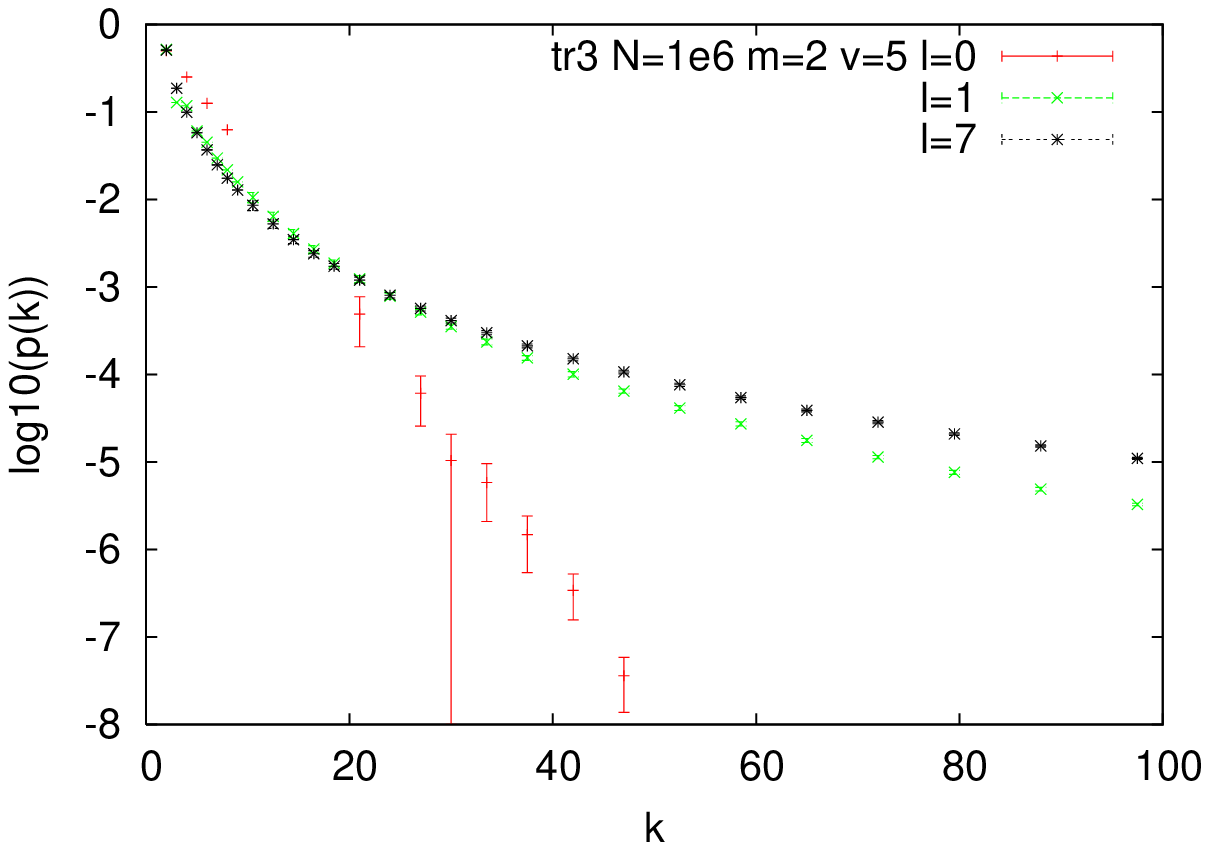}}
\\
 \scalebox{0.5}{\includegraphics{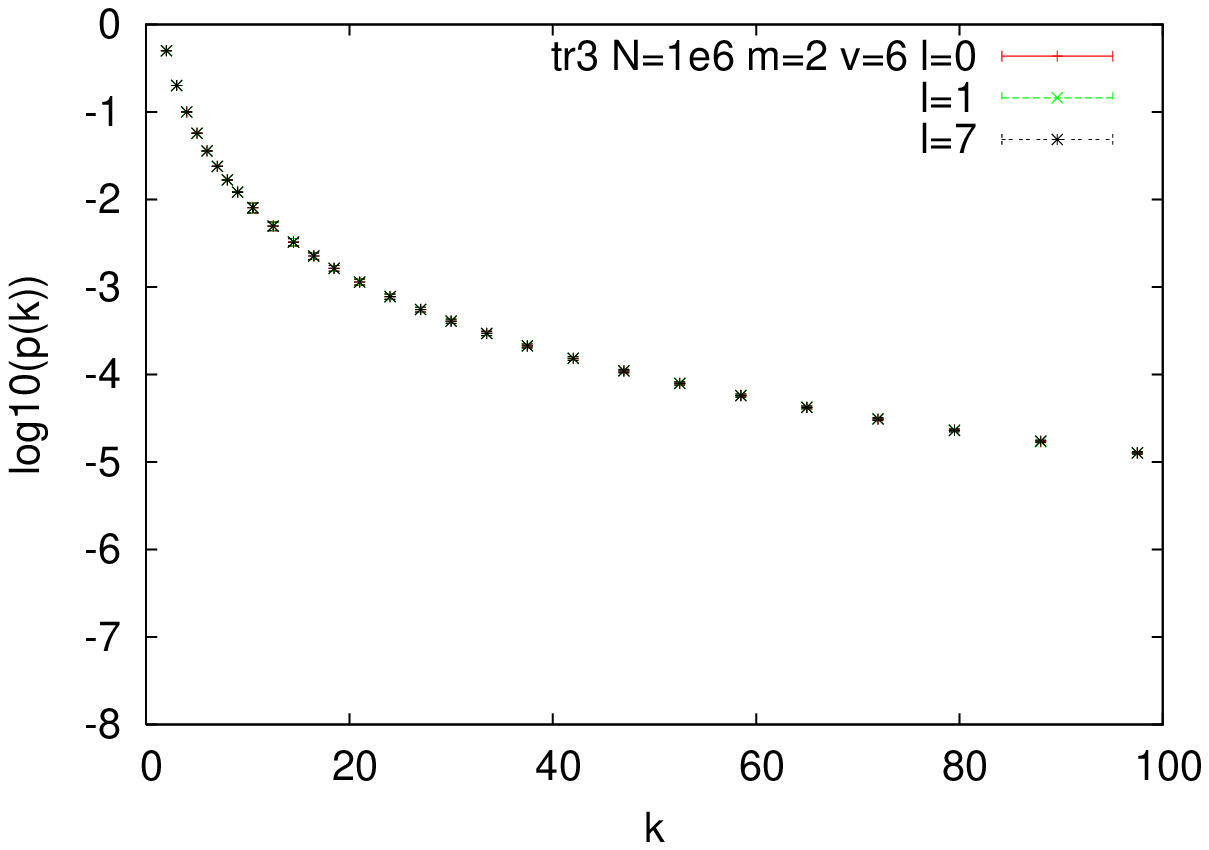}}
 \scalebox{0.5}{\includegraphics{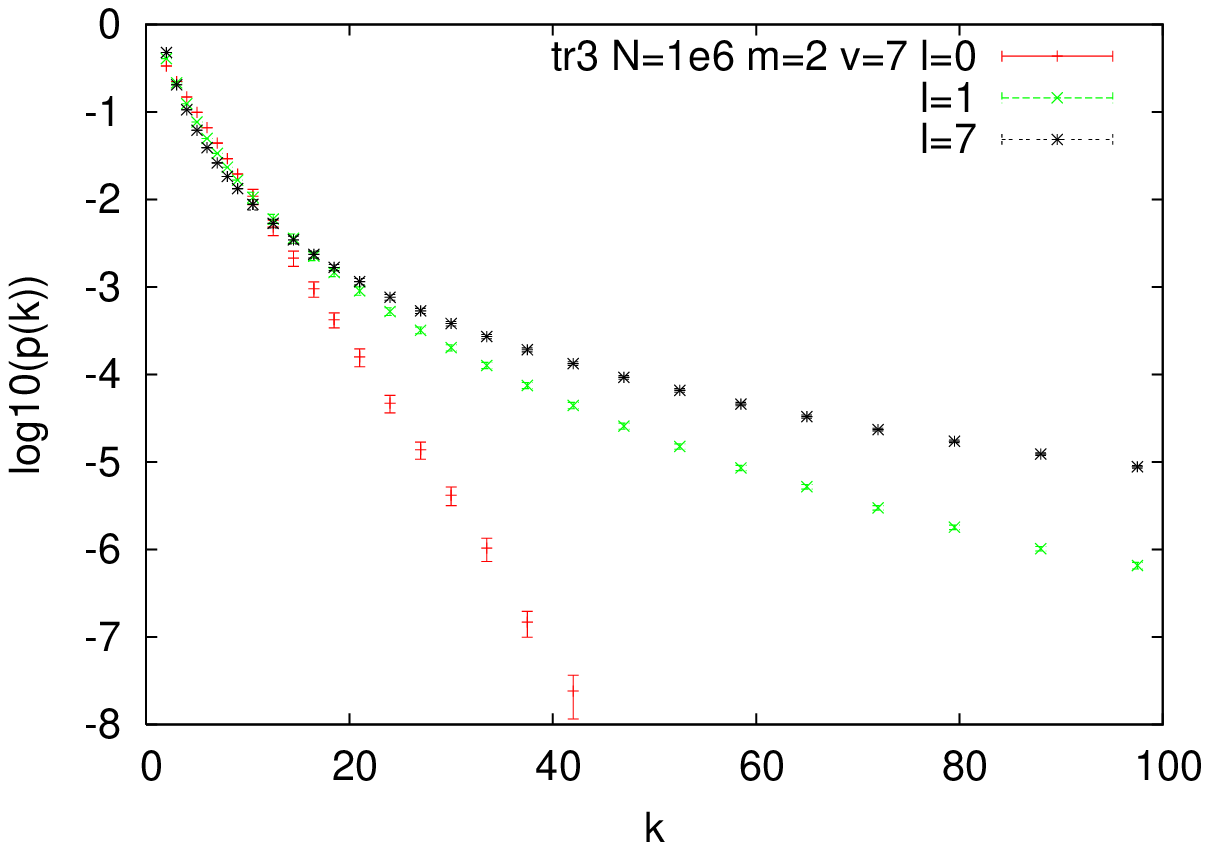}}
 \end{center}
\caption{Semi log plots of $\log_\mathrm{10}(n(k))$ vs $k$ for
algorithms $v=0$ top left $v=1$ top right and then in order down
to $v=7$ bottom right. All with one vertex ($\epsilon=1$) and two
edges ($m=2$) added at each time step and a total of $10^6$
vertices added. In each graph the results are shown for average
walk lengths (denoted by $s=l$) of 0,1 and 7 steps with data
averaged over 100 runs and the data are binned with bins chosen
such that $k_\mathrm{max}/k_\mathrm{min} \approx 1.1$. Multiple
edges are allowed here.}
 \label{ftr3n1e6s017v0_7all}
\end{figure}

Figure \ref{ftr3n1e6s017v0_7all} helps us to see the exponential
nature of the zero step walks when we start from a randomly chosen
vertex, i.e.\ $l=0$ walks.

\begin{figure}[!htb]
\begin{center}
 \scalebox{0.5}{\includegraphics{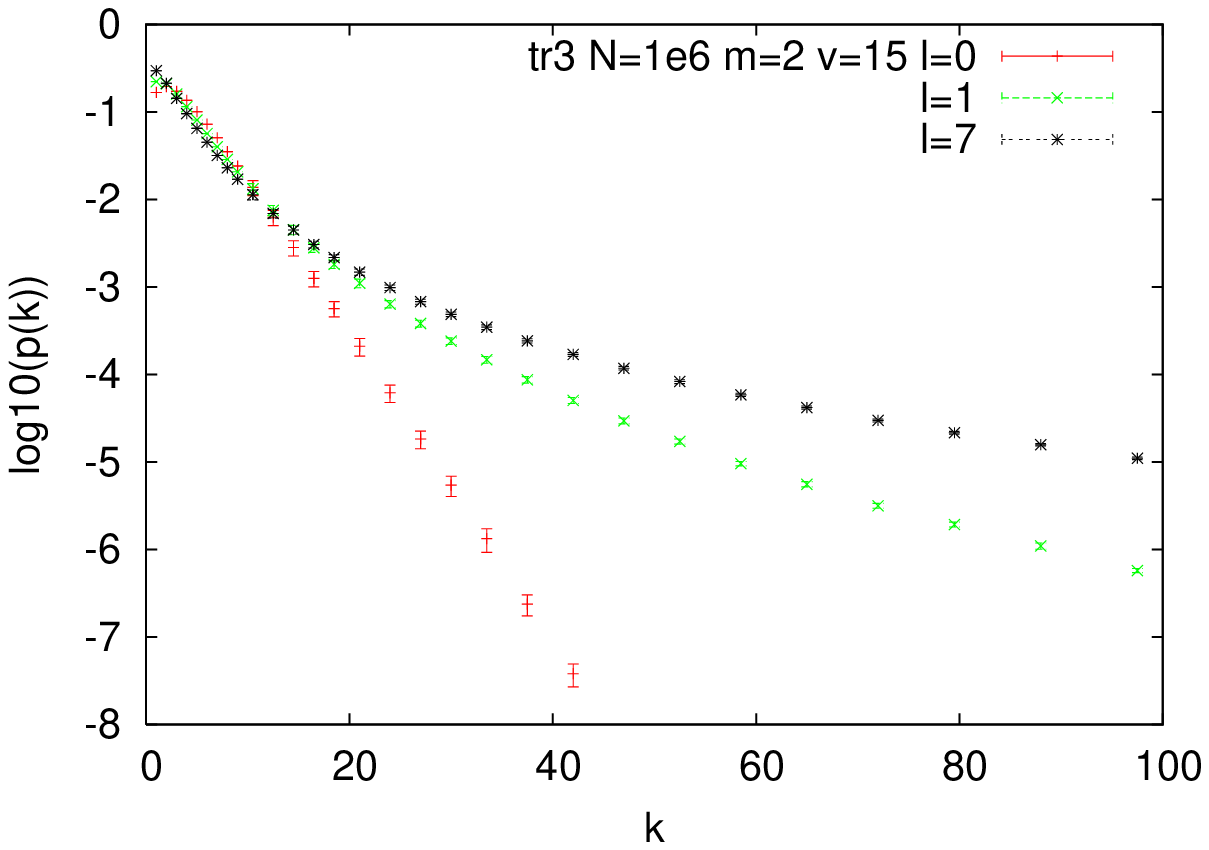}}
 \end{center}
\caption{Semi log plots of $\log_\mathrm{10}(n(k))$ vs $k$ for
algorithm $v=15$. With one vertex ($\epsilon=1$) and two edges
($m=2$) added at each time step and a total of $10^6$ vertices
added. In each graph the results are shown for average walk
lengths (denoted by $s=l$) of 0,1 and 7 steps with data averaged
over 100 runs and the data are binned with bins chosen such that
$k_\mathrm{max}/k_\mathrm{min} \approx 1.1$. Multiple edges are
allowed here.}
 \label{ftr3n1e6s017v15extra}
\end{figure}

\subsection{Finite Size Effects}

\begin{figure}[!htb]
\begin{center}
 \scalebox{0.6}{\includegraphics{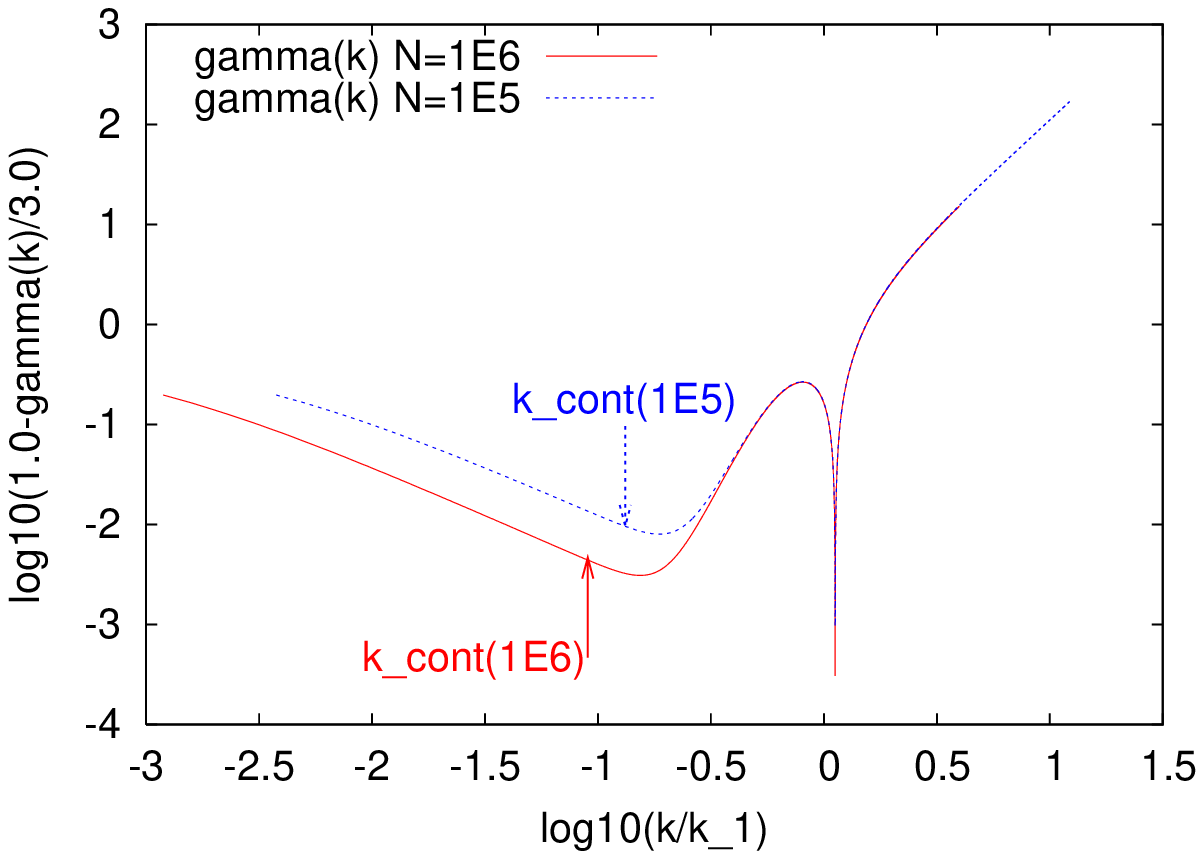}}
 \end{center}
\caption{The vertical axis is
$\log_\mathrm{10}(1-\gamma_\mathrm{eff}(k)/3.0)$, the log of the
fractional deviation of the mean field power
$\gamma_\mathrm{eff}(k)$ results from the large $N$ theoretical
prediction of a constant value of three.   The power
$\gamma_\mathrm{eff}(k)$ is obtained by fitting a power law to
neighbouring points in the mean field solution. This is plotted
against $\log_\mathrm{10} (k/k_1)$ where $k_\mathrm{1}$ should be
the degree of the largest vertex, i.e.\ the rank one vertex.one of
the scales implicit in any finite size sample.  Another scale,
$k_\mathrm{cont}$, which should be the end of the continuous
degree spectrum ($p(k_\mathrm{cont}) = 1/N$), is also indicated.}
 \label{fgammamffit}
\end{figure}

As discussed in the text, it is best to use data from as high a
scale as possible to avoid the finite size effects coming from the
small scales. We can use the effective local power
\tref{geffsmallk} as a good measure of the finite size deviations
by looking at the numerical (exact) solution to the mean field
equations, which are in turn an excellent approximation to pure
preferential attachment models (e.g.\ our random walk models with
a random edge start $\mathtt{(v \& 2)} = 0$). We can see that even
in this perfect case fitting a simple power law will not produce a
good result as figure \ref{fgammamffit} shows.

\begin{figure}[!htb]
\begin{center}
 \scalebox{0.6}{\includegraphics{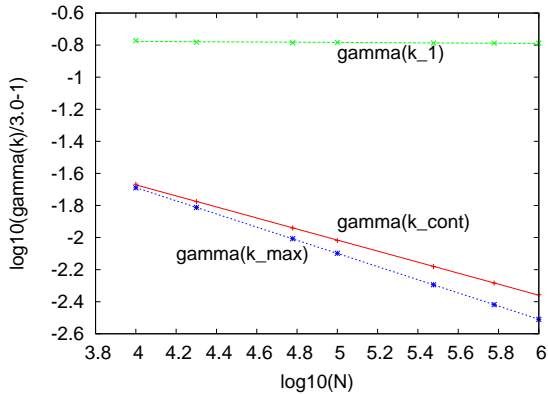}}
 \end{center}
\caption{Variation of different measures for the effective power
$\gamma_\mathrm{eff}(k)$ with $N$, for solutions to the mean field
equations with $\epsilon=1$ (pure preferential attachment) and
$m=2$. The straight lines are best fits to the data with slopes of
-0.34, -0.0065 and -0.41 for the fractional error in
$\gamma_\mathrm{eff}(k_1)$, $\gamma_\mathrm{eff}(k_\mathrm{cont})$
and $\gamma_\mathrm{eff}(k_\mathrm{max})$.}
 \label{mfgammavsN}
\end{figure}

We can consider the effective power $\gamma_\mathrm{eff}(k)$ at
the characteristic scales $k_\mathrm{cont}$ and $k_1$ and also at
$k_\mathrm{max}$ the degree with the largest power below $k_1$.
The fractional error between $\gamma_\mathrm{eff}(k)$ for finite
$N$ and infinite $N$ power value of three,
($(\gamma_\mathrm{eff}(k)/3.0 -1)$), is tending towards the large
N value as a power of $N$ as figure \ref{mfgammavsN}
shows\footnote{The results for the mean field model solution
\tref{ddmfinfsol} with $m=2$ are as follows. For $N=10^5$
$k_\mathrm{cont}= 105$ and $\gamma_\mathrm{eff}(k_\mathrm{cont})=
2.971$ while for $k_1= 796$ $\gamma_\mathrm{eff}(k_1)= 2.506$ with
a peak between these two values of
$\gamma_\mathrm{eff}(k_\mathrm{max})=2.976$ at
$k_\mathrm{max}=149$. For $N=10^6$ $k_\mathrm{cont}= 227$ and
$\gamma_\mathrm{eff}(k_\mathrm{cont})= 2.511$ while for $k_1=
2520$ $\gamma_\mathrm{eff}(k_1)= 2.987$ with a peak between these
two values of $\gamma_\mathrm{eff}(k_\mathrm{max})=2.991$ at
$k_\mathrm{max}=388$.}.
Its interesting to note though that for $k<k_1$, the power is
always below the large degree large $N$ value.  It is closest to
that theoretical value at $k_\mathrm{max}$ in a region a little
above $k_\mathrm{cont}$.

Good quality data is only available for $k \lesssim
k_\mathrm{cont}$ and the effective power obtained when fitting
these finite size but pure theoretical model results over a range
of degrees around $k_\mathrm{cont}$ is more likely to be 1\% (for
$N=10^6$) or 10\% (for $N=10^5$) below the large $N$ prediction.
Further, the data in figure \ref{fmfdatafit} was for one run of a
model which best represents the mean field equations and this
shows we must in practice expect larger deviations from the large
$N$ pure power law result \tref{ggensol}.

\subsection{Power law fits}

\begin{figure}[!htb]
\begin{center}
 \scalebox{0.6}{\includegraphics{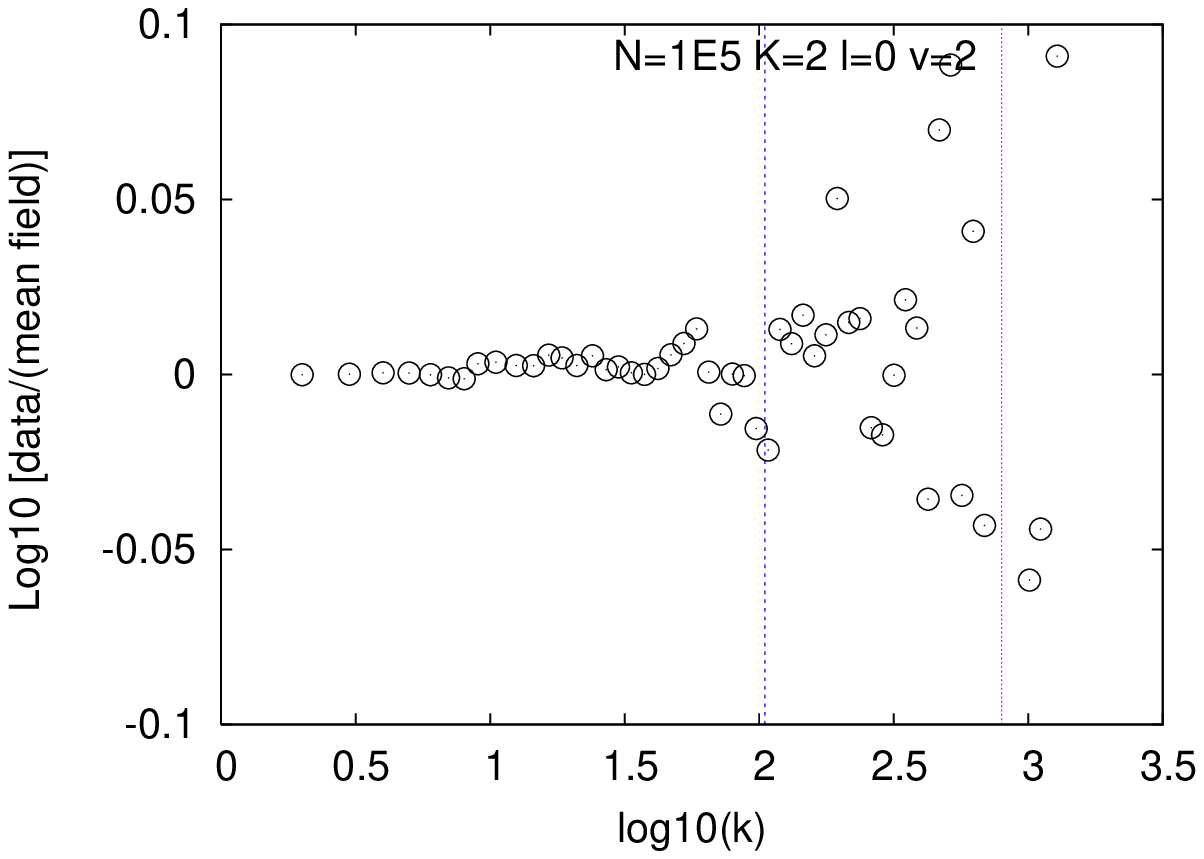}}
 \scalebox{0.6}{\includegraphics{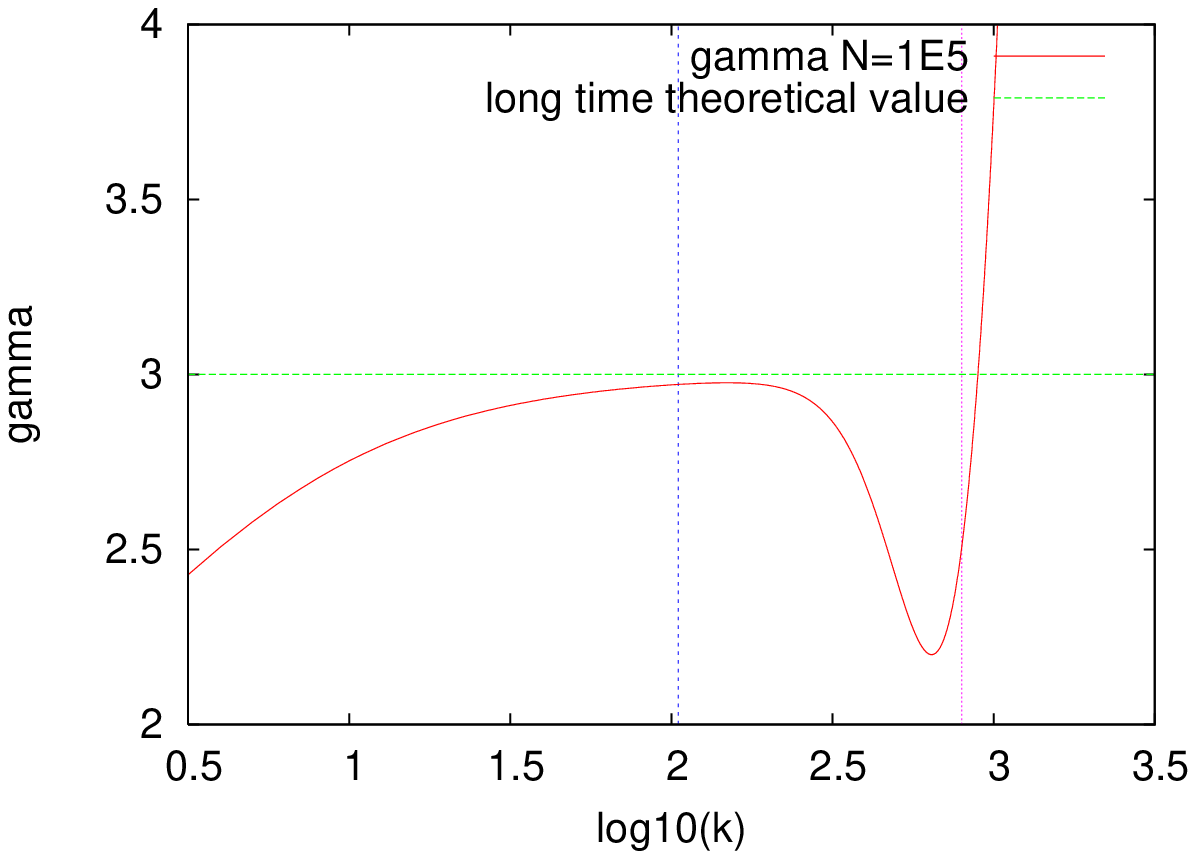}}
 \\
 \scalebox{0.6}{\includegraphics{mfn1e6k2fracerror.eps}}
 \scalebox{0.6}{\includegraphics{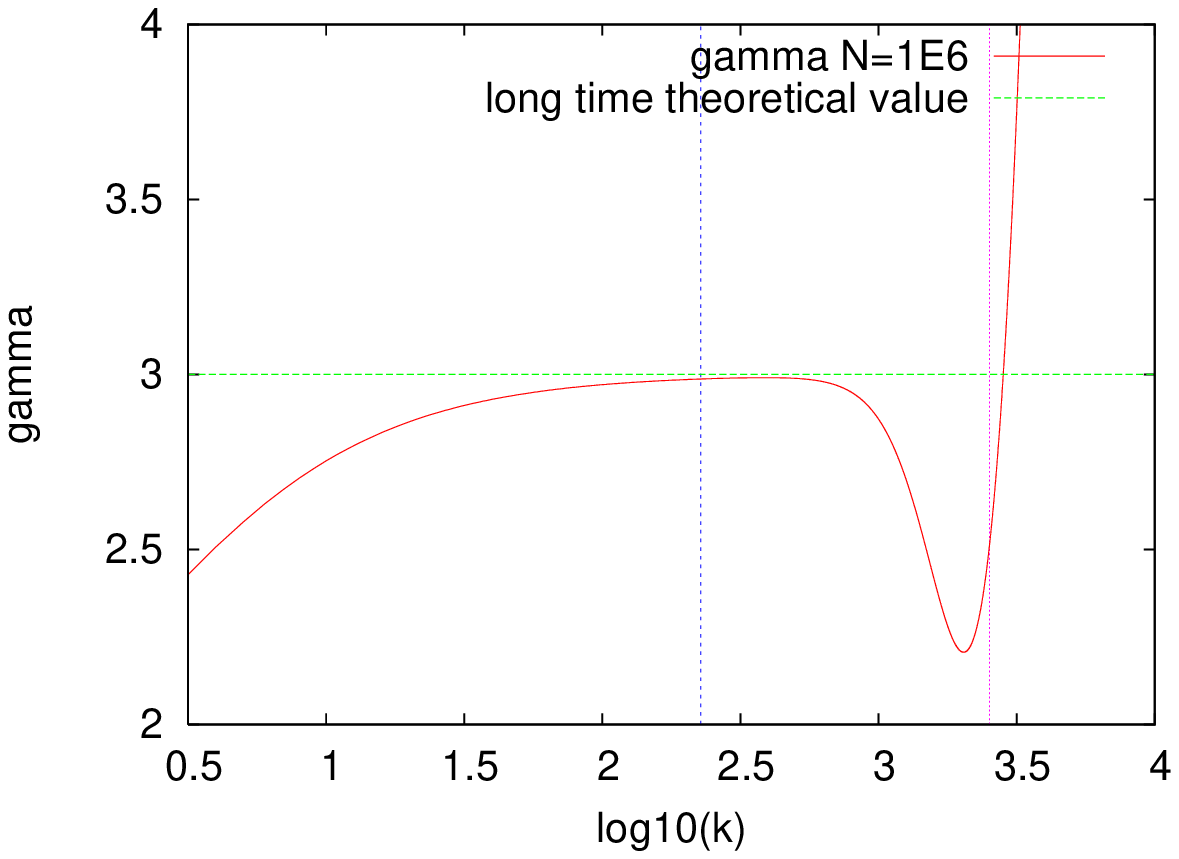}}
 \end{center}
\caption{On the left are shown plots of data (v=2, l=0, m=2,
averaged over 100 runs and log binned) normalised by the mean
field results. The power $\gamma$ obtained by fitting a power law
to neighbouring points in the mean field solution  with the
theoretical result $\gamma=3.0$ indicated. Top row for $N=10^5$
and bottom for $N=10^6$. The vertical lines show $k_\mathrm{cont}$
(on left) which should be the end of the continuous degree
spectrum ($p(k_\mathrm{cont}) = 1/N$), and $k_\mathrm{1}$ (on
right) which should be the degree of the largest vertex, i.e.\ the
rank one vertex.}
 \label{ftwver2e1e6}
\end{figure}

Further figures showing how the data fits the finite $N$ solutions
to the mean field equations well, but that these are not pure
power laws figure \ref{ftwver2e1e6}.


\subsection{Large degree scales}

\begin{figure}[!htb]
\begin{center}
 \scalebox{0.6}{\includegraphics{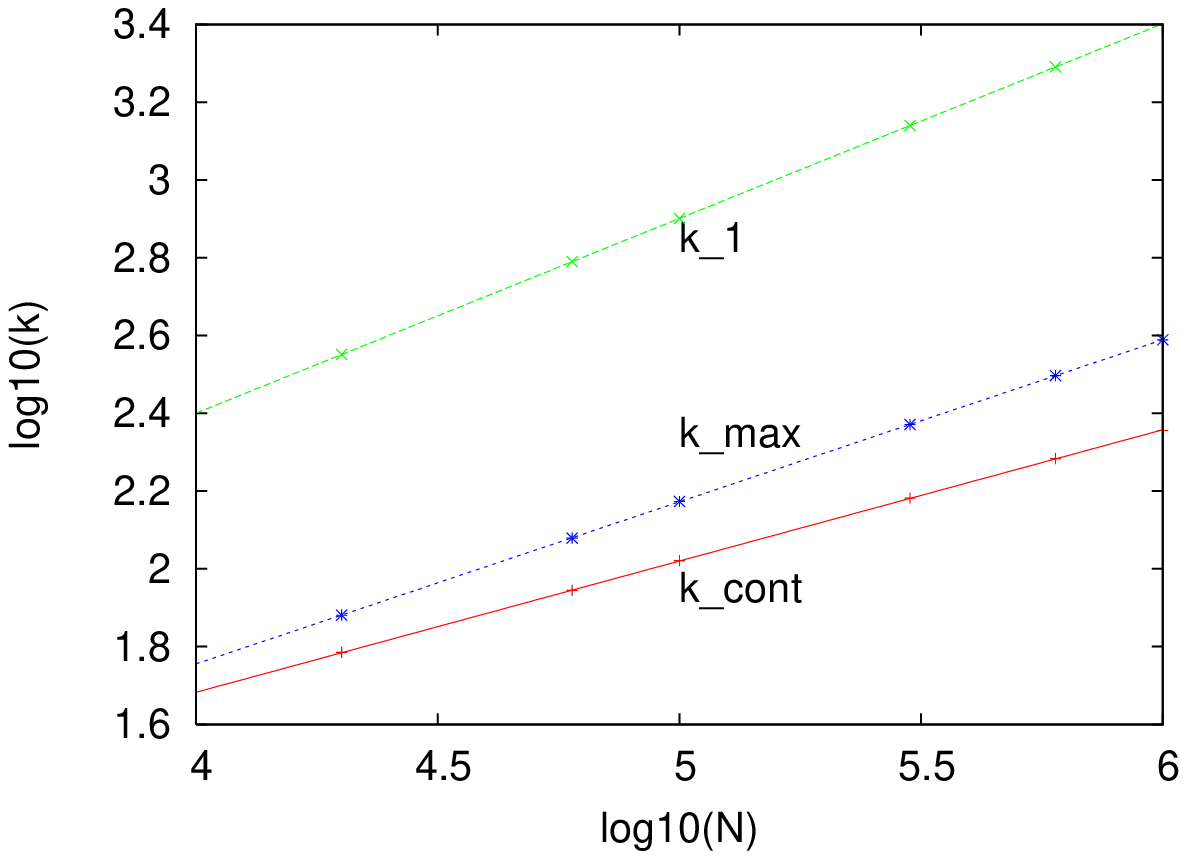}}
 \end{center}
\caption{Variation of different degree scales with $N$, for $m=2$,
 $\epsilon=1$. The points are solutions of the mean field equations for
$\epsilon=1$, $m=2$ so $r=2$. The straight lines are best line
fits to the data with slopes of 0.337, -0.501 and -0.417 for the
fractional error in $\gamma_\mathrm{eff}(k_1)$,
$\gamma_\mathrm{eff}(k_\mathrm{cont})$ and
$\gamma_\mathrm{eff}(k_\mathrm{max})$.}
 \label{mfkvsN}
\end{figure}

It is useful to use the characteristic degree scales of
$k_\mathrm{cont}$ \tref{kcontdef} and $k_1$ \tref{k1def} which
mark the region where the largest $k$ values can be extracted from
the data.  Since $\gamma=3$ is the infinite $N$ solution for the
parameter values used here ($\epsilon=2$, $p_v=1$), the degree
scales might be expected to vary as $k_\mathrm{cont} \propto
N^{1/3}$ and $k_1 \propto N^{1/2}$ and indeed we see this scaling
in figure \ref{mfkvsN}.

\subsection{Distance measures}

\begin{figure}[htb]
\begin{center}
 \scalebox{0.5}{\includegraphics{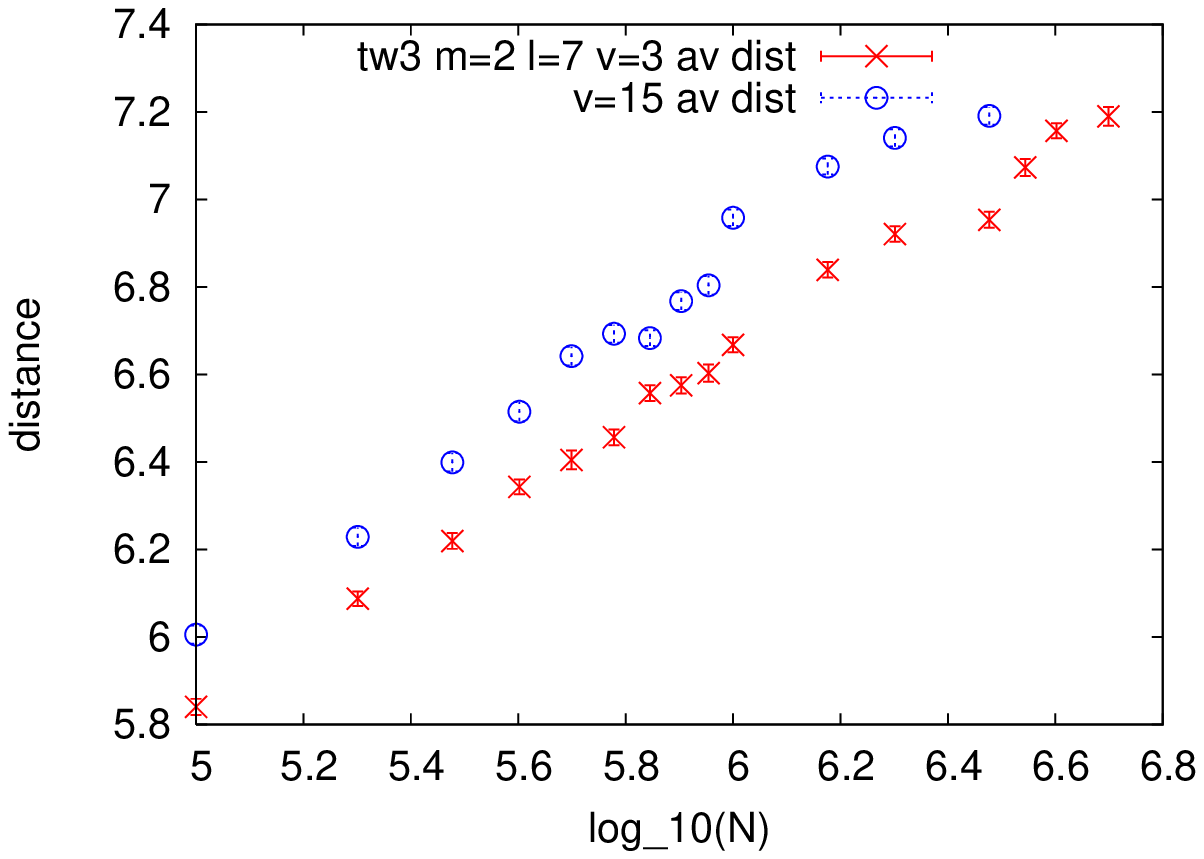}}
 \scalebox{0.5}{\includegraphics{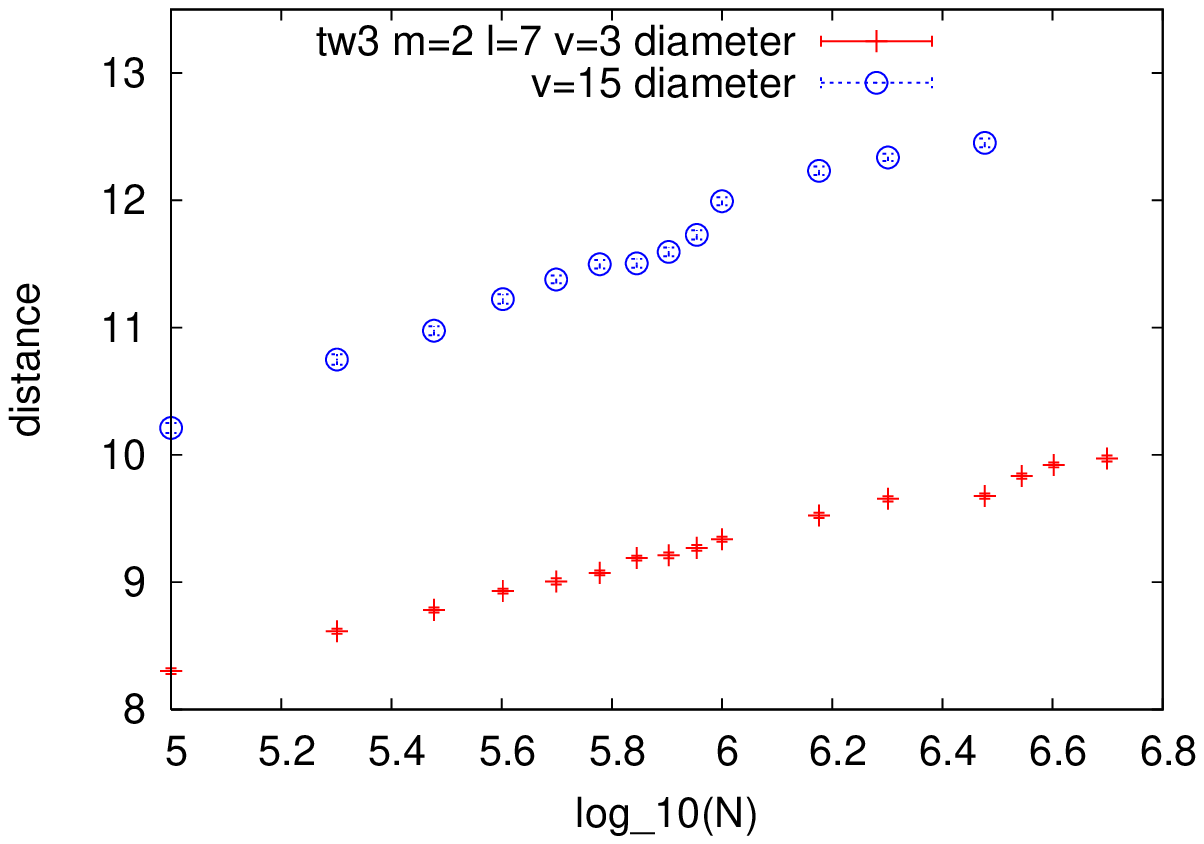}}
 \end{center}
\caption{Average shortest distances and diameters for different
total numbers of vertices $N$, with the average degree held fixed
($K=2$).  The error bars on data points are drawn but are
comparable with the size of the symbol. The $v=3$ data are for 100
runs a new random walk starting for every edge added and of fixed
length $l=7$.  The second example allows a variable number of
steps in the random walk and a variable number of vertices added
at each step but keep the averages the same as before ($v=15$).}
 \label{fdistvaryN}
\end{figure}

We can look at the diameter and the average shortest distance
between points for different algorithms, see figure
\ref{fdistvaryN}.


\begin{thebibliography}{99}



\bibitem{AB} R.Albert and A.-L.Barab\'asi,
 \tpretitle{Statistical Mechanics of Complex Networks}
  Rev.Mod.Phys.\ \vol{74} (2002) 47
 [\eprint{cond-mat/0106096}].

\bibitem{DM01}
  S.N.Dorogovtsev and J.F.F.Mendes,
 \tpaptitle{Evolution of networks}
 Adv.Phys.\ \vol{51} (2002) 1079
 [\eprint{cond-mat/0106144}].

\bibitem{Watts03}
 D.J.Watts,
 \tbktitle{Six Degrees: The science of a Connected Age}
 (Heinemann, London, 2003).

\bibitem{Newman03}
 M.E.J. Newman, \tpaptitle{The structure and function of complex networks}
 SIAM Review \vol{45} (2003) 167.

\bibitem{TSE04}
 T.S.Evans,
 \tpaptitle{Complex Networks}
 Contemporary Physics \vol{45} (2004) 455
 [\eprint{cond-mat/0405123}].

\bibitem{ER59} P.Erd\H{o}s and A.R\'enyi,
 \tpaptitle{On Random Graphs I}
 Publicationes Mathematicae (Debrecen) \vol{6} (1959) 290.

\bibitem{Mit04}
 M.Mitzenmacher,
 \tpaptitle{A Brief History of Generative Models for Power Law and Lognormal
 Distributions}
 Internet Mathematics \vol{1} (2004) 226
 [\texttt{http://www.internetmathematics.org/}]

\bibitem{Simon55}
 H.A.Simon,
 \tpaptitle{On a Class of Skew Distribution Functions}
 Biometrika \vol{42} (1955) 425.

\bibitem{BA}
 A.-L.Barab\'{a}si and R.Albert,
 \tarttitle{Emergence of scaling in random networks}
  Science \vol{286} (1999) 173.

 \bibitem{KRL}
 P.L.Krapivsky, S.Redner and F.Leyvraz,
 \tpretitle{Connectivity of Growing Random Networks}
 Phys.Rev.Lett. \vol{85} (2000) 4629
 [\eprint{cond-mat/0005139}].

\bibitem{Vaz00}
  A.V\'azquez,
  \tpretitle{Knowing a network by walking on it: emergence of
  scaling}
  \eprint{cond-mat/0006132}.

\bibitem{Vaz02}
  A.V\'azquez,
  \tpaptitle{Growing networks with local rules: preferential attachment, clustering hierarchy
and degree correlations}
 Phys. Rev.E \vol{67} (2003)  056104
 [\eprint{cond-mat/0211528}].

\bibitem{KR02} P.L.Krapivsky and S.Redner,
 \tarttitle{Finiteness and Fluctuations in Growing Networks}
 J.Phys.A \vol{35} 9517 (2002)
 [\eprint{cond-mat/0207107}].

\bibitem{Chung02}
 F. Chung, L. Lu, T.G. Dewey, and D.J. Galas,
 \tpaptitle{Duplication Models for Biological Networks}
 J.Computational Biology \vol{10} (2003) 677.

\bibitem{SK04}
 J.Saram\"aki and K.Kaski,
 \tpretitle{Scale-Free Networks Generated by Random Walkers}
 \eprint{cond-mat/0404088}.

\bibitem{Kla}
 S.Klauke,
 \tpretitle{Scale-free Networks and the Random Walk Algorithm}
 Imperial College BSc project report, April 2002.


\bibitem{KRR} P.L.Krapivsky, G.J.Rogers and S.Redner,
  \tpaptitle{Degree Distributions of Growing Networks}
  Phys.Rev.Lett. \vol{86} (2001) 5401
  [\eprint{cond-mat/0012181}].

\bibitem{WDN02}
 D.J.Watts, P.S.Dodds and M.E.J.Newman,
 \tarttitle{Identity and Search in Social Networks}
 Science \vol{296} (2002) 1302-1305.

\bibitem{BAJ}
 A.-L.Barab\'asi, R.Albert and H.Jeong,
 \tpaptitle{Mean-field theory for scale-free random networks}
  Physica A \vol{272} (1999) 173.
  [\eprint{cond-mat/9907068}].

\bibitem{DMS00}
  S.N.Dorogovtsev, J.F.F.Mendes and A.N.Samukhin,
 \tpaptitle{Structure of Growing Networks With Preferential Linking}
 Phys.Rev.Lett.\ \vol{85} (2000) 4633-4636
 [\eprint{cond-mat/0004434}].



\bibitem{CNSW}
 D.S.Callaway. M.E.J.Newman, S.H.Strogatz and D.J.Watts,
 \tpaptitle{Network Robustness and Fragility: Percolation on
 Random Graphs}
 Phys.Rev.Lett. \vol{85} (2000) 5648.

\bibitem{KR} P.L.Krapivsky and S.Redner,
 \tarttitle{Organisation of Growing Random Networks}
 Phys.Rev.E \vol{63} (2001) 066123-1
 [\eprint{cond-mat/0011094}].


\bibitem{DMS01}
 S.N.Dorogovtsev, J.F.F.Mendes and A.M.Samukhin,
 \tpaptitle{Size-dependent degree distribution of a scale-free
 growing network}
 Phys.Rev.E \vol{63} (2001) 062101
 [\eprint{cond-mat/0011115}].

\bibitem{KK00}
 L.Kullmann and J.Kert\'esz,
 \tpaptitle{Preferencial growth: exact solution of the time
 dependent distributions}
 Phys.Rev.E \vol{63} (2001) 051112
 [\eprint{cond-mat/0012410}].

\bibitem{ZM00}
 D.H.Zanette and S.C.Manrubia,
 \tpretitle{Vertical transmission of culture and the distribution of family
names}
 \eprint{nlin.AO/0009046}.

\bibitem{YookWeightedSF}
S.H. Yook, H. Jeong, A.-L. Barab\'{a}si and Y. Tu,
\tpaptitle{Weighted evolving networks},
 Phys.Rev.Lett.\vol{86} (2001) 5835.


 \bibitem{BBV04b}  A.Barrat, M.Bart\'{e}lemy, R.Pastor-Satorras and A.Vespignani,
\tpaptitle{The architecture of complex weighted networks}
 Proc.Natl.Acad.Sci USA \vol{101} (2004) 3747

\bibitem{BBV} A.Barrat, M.Bart\'{e}lemy and A.Vespignani,
 \tpaptitle{Weighted evolving networks: coupling topology and
 weights dynamics}
 Phys.Rev.Lett \vol{92} (2004) 228701
 [\eprint{cond-mat/0401057}].


\bibitem{NewmanWeighted2004}
 M.E.J. Newman,
 \tpaptitle{Analysis of weighted networks}
 Phys.Rev.E \vol{70} (2004) 056131
 [\eprint{cond-mat/0407503}].


\bibitem{Onnela2004}
 J.-P. Onnela, J. Saram\"{a}ki, J. Kert\'{e}sz, and K. Kaski,
 \tpaptitle{Intensity and coherence of motifs in weighted complex networks}
 Phys.Rev.E, in press (2005)
 \eprint{cond-mat/0408629}.




\end{thebibliography}
\end{document}